\documentclass[aps,rmp,superscriptaddress,showpacs,floatfix,amsmath,amssymb,notitlepage,nofootinbib]{revtex4-1}
\usepackage{graphicx}
\usepackage[pdftex, pdfstartview = {FitH}]{hyperref}
\usepackage[normalem]{ulem}

\DeclareMathOperator{\trace}{Tr}
\newcommand{\rv}{ {\boldsymbol{r}} }
\newcommand{\xv}{ {\boldsymbol{x}} }
\newcommand{\sv}{ {\boldsymbol{s}} }
\newcommand{\jv}{ {\boldsymbol{\jmath}} }
\newcommand{\vv}{ {\boldsymbol{v}} }
\newcommand{\Lv}{ {\underline{\boldsymbol{L}}} }
\newcommand{\nablav}{ {\boldsymbol{\nabla}} }
\newcommand{\be}[0]{ \begin{equation} }
\newcommand{\ee}[0]{ \end{equation} }
\newcommand{\ber}[0]{ \begin{eqnarray} }
\newcommand{\eer}[0]{ \end{eqnarray} }
\newcommand{\nn}[0]{ \nonumber \\ }

\renewcommand{\vr}[0]{ \boldsymbol{ r } }
\newcommand{\mat}[1]{ \underline{\boldsymbol #1} }
\newcommand{\vrp}[0]{ \vr^{\prime} }
\newcommand{\Ev}[0]{ \boldsymbol{E} }
\newcommand{\Av}{{ \bf A} }
\newcommand{\ks}[0]{ \mathrm{s} }
\newcommand{\xc}[0]{ \mathrm{xc} }
\newcommand{\Hartree}[0]{ \mathrm{H} }
\newcommand{\Hxc}[0]{ \mathrm{Hxc} }
\newcommand{\eq}[0]{ \mathrm{eq} }

\newcommand{\adia}[0]{ \mathrm{adia} }
\newcommand{\ALDA}[0]{ \mathrm{ALDA} }
\newcommand{\dyn}[0]{ \mathrm{dyn} }
\newcommand{\thermal}[0]{ \mathrm{th} }
\newcommand{\visc}[0]{ \mathrm{visc} }
\newcommand{\eff}[0]{ \mathrm{eff} }
\newcommand{\fermi}[0]{ \mathrm{F} }
\newcommand{\kB}[0]{ k_{\mathrm{B}} }
\newcommand{\fder}[2]{ \frac{\delta#1}{\delta#2} }

\newcommand{\intc}[0]{ \int_{\mathcal{C}} }
\newcommand{\intcdef}[0]{ \int_{\mathcal{C}} \tfrac{d t(\tau)}{d \tau} \mathrm{d}\tau }
\newcommand{\ind}[2]{ \int\!\!\mathrm{d}^{#1}\!#2 \; }
\newcommand{\indll}[3]{ \int_{#2}^{#3}\!\!\mathrm{d}#1 \; }

\begin{document}

\title{ Functional theories of thermoelectric phenomena }

\author{F. G. Eich}
\email[]{florian.eich@mpsd.mpg.de}
\affiliation{Max-Planck-Institute for the Structure and Dynamics of Matter, Luruper Chaussee 149, D-22761 Hamburg, Germany}
\affiliation{Department of Physics, University of Missouri-Columbia, Columbia, Missouri 65211, USA}

\author{M. \surname{Di Ventra}}
\affiliation{University of California at San Diego, La Jolla, California 92093, USA}

\author{G. Vignale}
\affiliation{Department of Physics, University of Missouri-Columbia, Columbia, Missouri 65211, USA}

\begin{abstract}
  We review the progress that has been recently made  in the application of time-dependent density functional theory to thermoelectric phenomena.  As the field is very young, we emphasize open problems and fundamental issues. We begin by introducing the formal structure of \emph{thermal density functional theory}, a density functional theory with two basic variables -- the density and the energy density -- and two conjugate fields -- the ordinary scalar potential and Luttinger's thermomechanical potential.   The static version of this theory is contrasted with the familiar finite-temperature density functional theory, in which only the density is a variable.  We then proceed to constructing the full time-dependent non equilibrium theory, including the practically important Kohn-Sham equations that go with it.  The theory is shown to recover standard results of the Landauer theory for thermal transport in the steady state, while showing greater flexibility by allowing a description of fast thermal response, temperature oscillations and related phenomena. Several results are presented here for the first time, i.e., the proof of invertibility of the thermal response function in the linear regime, the full expression of the thermal currents in the presence of Luttinger's thermomechanical potential, an explicit prescription for the evaluation of the Kohn-Sham potentials in the adiabatic local density approximation, a detailed discussion of the leading dissipative corrections to the adiabatic local density approximation and the thermal corrections to the resistivity that follow from it.
\end{abstract} 

\date{\today}

\maketitle

\tableofcontents

\section{Introduction}

Density functional methods occupy a central position in the landscape of modern computational materials theory~\cite{DreizlerGross:90,GiulianiVignale:05,MarquesRubio:12,Ullrich:12}. Not only do they provide an indispensable tool for the calculation of equilibrium properties of materials, but they are also being widely used for the calculation of transport properties at all length scales~\cite{Ullrich:12,DiVentra:08}. Key to the success of these methods is their ability to incorporate important many-body effects, arising from electron-electron interaction, in an intuitive and computationally affordable scheme. In spite of all this, little attention has been paid so far to the possibility of applying density functional methods directly to the study of thermoelectric phenomena. These are transport phenomena, such as the Seebeck and the Peltier effect, in which electronic degrees of freedom are involved in an essential manner, with important contributions from electron-phonon interactions~\cite{NolasGoldsmid:01,DubiDiVentra:11}. Therefore, they seem natural candidates for a time-dependent non equilibrium density functional treatment. Beyond its immediate practical value, such a treatment would forge a link between density functional theory -- a formally exact theory valid at all length scales -- and classical hydrodynamics, a theory of locally conserved particle, momentum and energy energy density, which is valid at long wavelengths and long time scales~\cite{LandauLifshitz-6,AndreevSpivak:11}. Such a link is not complete at present, due to the absence of thermal and thermoelectric couplings in density functional theory.

In this article, we review the progress that has recently been made in the application of non equilibrium density functional theory to thermoelectric phenomena~\cite{EichVignale:14a}. Since this line of research is a very recent development, we focus specifically on open problems and fundamental issues. We start by introducing the basic variable of the time-dependent thermal density functional theory, namely, the energy density, and discuss the problems associated with its non-unique definition. Next, following a classic paper by Luttinger~\cite{Luttinger:64a}, we introduce the mechanical field conjugate to the energy density, which we call ``thermomechanical potential''. This thermomechanical potential serves as mechanical proxy for local temperature variations~\cite{Shastry:09}. We discuss how the results of the standard Landauer theory~\cite{Landauer:57,Landauer:89}, for mesoscopic thermal transport can be reproduced in the framework of the thermomechanical potential approach, while the latter exhibits higher flexibility in the nonlinear regime~\cite{EichVignale:14b}. We conclude by introducing the generalization of the thermomechanical potential to the ``thermal vector potential'' that couples to thermal currents~\cite{Tatara:15a}.

We then proceed to develop the formalism of the new functional theory from a generalized stationary action principle and introduce the Kohn-Sham reference system and the formal construction of its one-particle effective potentials~\cite{EichVignale:14a}. This formalism is suitable for the calculation of electrical and thermal currents, as well as particle and energy densities. At this point we present a comparison between the thermal density functional theory and alternative approaches that can potentially be applied to the same class of problems, e.g., the stochastic density functional theory for open systems~\cite{DagostaDiVentra:07,BieleDagosta:12,DagostaDiVentra:13}, and the time-dependent thermal transport theory~\cite{BieleRubio:15}, and discuss ways in which these different theories could complement each other.

The last part of the review is concerned with the practical issue of constructing an explicit functional for thermal density functional theory: we examine the adiabatic local density approximation and the leading dissipative corrections to it in the light of recent numerical calculations of the free energy density~\cite{BrownCeperley:13a,KarasievTrickey:14} and thermal transport coefficient of the homogeneous electron gas at finite temperatures.  Lastly, we present a few applications of the theory which are relevant to experiments such as the existence of thermal dynamical corrections to the electrical resistance of conductors~\cite{SaiDiVentra:05,KoentoppEvers:06,VignaleDiVentra:09,KurthStefanucci:13}, current- and thermal-current-induced quantum oscillations in the local temperature~\cite{BergfieldDiVentra:15,EichVignale:16}, transient flow of charge and energy and dynamical temperature waves in mesoscopic systems~\cite{EichVignale:16}.

\section{Luttinger's thermomechanical potential} \label{SEC:TMpotential}

Hydrodynamics is one of the most versatile and successful methods ever introduced to describe the dynamics of strongly interacting many-particle systems.  In its full fledged form the hydrodynamic theory consists of three equations~\cite{HuangTransport:87,PuffGillis:68}: the continuity equation, which connects the time derivative of the particle density, $n(\rv)$, to the divergence of the particle current density $\jv_n(\rv)$ (proportional to the momentum density); the Euler equation, which expresses the time-derivative of the particle current density in terms of the divergence of an internal stress tensor $\mat{\sigma}(\rv)$, plus external volume forces; and the heat transport equation, which expresses the time derivative of the energy density, $h(\rv)$ in terms of the divergence of the energy current density $\jv_h(\rv)$.  Essential to the closure of the scheme is the existence of approximate linear relations that allow us to express the stress tensor (alias momentum current) and the heat current, $\jv_q(\rv)=\jv_h(\rv)-\mu\jv_n(\rv)$,  in terms of the gradients of local velocity and temperature fields, which in turn are determined by the particle current density and energy density. There is a very good reason why hydrodynamics employs the densities of particles, momentum, and energy, and no more.  In the limit of slow spatial variation these are quasi-conserved quantities and therefore their time evolution is very slow.  Provided the external potentials are slowly varying on the microscopic scale, the evolution of the quasi conserved quantities is sufficient to characterize the dynamics of the system, especially so when the particle-particle interactions are strong and frequent, resulting in small deviations from local thermal equilibrium.

Electrons in solid state devices rarely satisfy the conditions for the applicability of hydrodynamics: the external potentials arising from nuclei, impurities, lattice vibrations, are usually not slowly varying on the microscopic scale. Further, the description of the particles is necessarily quantum mechanical.  Nevertheless, a powerful many-body formalism known as time-dependent density functional theory (TDDFT)~\cite{RungeGross:84,Tokatly:05a,Tokatly:05b,Ullrich:12} allows us to derive closed equations of motion for the particle density and the current density, similar to the hydrodynamic equations, but  valid, in principle, at all length scales.  The foundation of the theory is the Runge-Gross theorem~\cite{RungeGross:84}, according to which the quantum stress tensor is uniquely determined by the density and the initial state of the system.  In practice, however, the dependence of the stress tensor on the underlying densities is strongly nonlocal--both, in space and time--and needs to be a subject of  drastic approximations.  Further, the Fermi statistics of the electrons forces in many cases an orbital representation of the densities, the so-called Kohn-Sham representation~\cite{KohnSham:65}, which tends to obscure the hydrodynamic structure of the theory. In spite of these limitations, TDDFT and its companion time-dependent current-density functional theory (TDCDFT)~\cite{Vignale:05} have grown to be widely used approximation methods in the study of electronic dynamics, especially optical excitations and electric transport, at the nanoscale~\cite{MarquesRubio:12}.

The topic of this review is the progress that has recently been made in extending the TDDFT so that it becomes capable to deal with thermoelectric phenomena, such as heat transport, local heating in current-carrying systems, temperature-voltage conversion in nanoscale electronic devices.  Our focus is on phenomena that depend essentially on electronic degrees of freedom.  There is, of course, significant heat transport in electrically insulating materials, where the excitations responsible for the transport are phonons or magnons.  As a first step we concentrate here on the problem of calculating heat currents in electric, non magnetic conductors, where heat is primarily carried by electrons in the vicinity of the Fermi surface. Parts of our system may be in contact with external reservoirs, which enforce a local temperature.  Other parts may be allowed to ``float" so as to adjust their temperature -- assuming that such a notion still makes sense -- to the value that is most compatible with the microscopic state of the system. In Sec.\ \ref{SEC:TDthermalTransport} we use these floating probe leads to compute the local temperature profile in a conducting nano wire.

\subsection{Thermal energy density} \label{SEC:localEnergy}

A major difficulty in carrying out such a program is that quantities such as temperature and entropy have a statistical significance, but not a clear mechanical one.  TDDFT, on the other hand is designed as a quantum mechanical theory, in which the quantum states evolve deterministically starting from some initial equilibrium ensemble, according to the equations of quantum mechanics (for a stochastic approach, in which the electrons are allowed to interact with a ``thermal bath", see section~\ref{SEC:StochasticDFT}). The unperturbed Hamiltonian for such a system is
\be
\hat H_0 = \hat T_0 +\hat V_0+\hat U_0 
\ee
where
\be \label{KineticEnergy}
\hat T_0 = \ind{3}{r}  \frac{\hbar^2}{2m} \big[ \nablav \hat \Phi^\dagger(\rv) \big] \cdot \big[ \nablav \hat \Phi(\rv) \big] 
\ee
is the kinetic energy,
\be
\hat V_0 = \ind{3}{r}~ v(\rv) \Phi^\dagger(\rv) \hat \Phi(\rv)
\ee
is the external (static) potential energy, and
\be
\hat U_0 = \frac{1}{2} \ind{3}{r} \ind{3}{s} \hat \Phi^\dagger(\rv) \hat \Phi^\dagger(\sv)w(|\rv-\sv|) \hat \Phi^\dagger(\sv) \hat \Phi(\rv)
\ee
is the electron-electron interaction energy.  The basic variables of standard TD(C)DFT are the particle density
\be
\hat n(\rv) = \hat \Phi^\dagger(\rv) \hat \Phi(\rv) 
\ee
and the particle current density
\be\label{CurrentDensity}
\hat \jv_{n,0} (\rv) = 
\frac{i\hbar}{2m} \left( \big[\nablav \Phi^\dagger(\rv) \big] \hat \Phi(\rv)-\hat \Phi^\dagger(\rv) \big[\nablav \hat \Phi(\rv) \big] \right)~.
\ee
The corresponding expectation values are denoted by the same symbols without the hat, i.e., $n(\rv)$ and $\jv_{n,0}(\rv)$.  All these quantities have a clear mechanical significance and are unambiguous insofar as they are the sources of a measurable electromagnetic field.  

To introduce thermal effects, the natural local variable would be the entropy density.  According to equilibrium statistical mechanics the entropy is the expectation value
\be\label{Entropy1}
S_0= \beta \left\langle \hat H_0 - \mu \hat N -\Omega_0 \right\rangle
\ee
where $\beta \equiv \frac{1}{k_BT}$ is the inverse equilibrium temperature,  $\mu$ is the chemical potential, $\hat N = \ind{3}{r}~\hat n(\rv)$ is the total number operator,  and  $\Omega_0$ is the grand thermodynamic potential defined as
\be
e^{-\beta\Omega_0}= {\rm Tr}~e^{-\beta(\hat H_0-\mu \hat N)}\,.
\ee
The equilibrium expectation value of any operator $\hat A$ is 
\be
\langle \hat A \rangle = {\rm Tr} \left[ \hat A e^{-\beta(\hat H_0-\mu\hat N -\Omega_0)}\right] ~.
\ee
Since $\Omega_0$ is just a number, which commutes with the Hamiltonian,  Eq.~(\ref{Entropy1}) strongly suggests that the operator
\be\label{ThermalHamiltonian}
\hat Q_0 \equiv \hat H_0 - \mu \hat N_0
\ee
 is the mechanical operator that corresponds to the entropy (times the temperature).   In fact, this is not quite true, due to the presence of the $\Omega_0$ term in Eq.~(\ref{Entropy1}).  Nevertheless, the role played by this quantity in the theory of thermal transport is so central that we will from now on refer to $\hat Q_0$ as the ``thermal Hamiltonian"   and we introduce the associated thermal energy density operator $\hat q_0(\rv)$, in the following manner
\begin{subequations} \label{EnergyDensities}
  \begin{align} 
    \hat q_0(\rv) & = \hat{h}_0(\rv)  - \mu\hat n(\rv) ~, \label{ThermalEnergyDensity} \\ 
    \hat{h}_0(\rv) & = \hat t_0(\rv) + \hat v_0(\rv)+ \hat u_0 (\rv) ~, \label{EnergyDensity} \\
    \hat t_0 (\rv) & = \frac{\hbar^2}{2m} \big[ \nablav \hat \Phi^\dagger(\rv) \big] \cdot \big[ \nablav \hat \Phi(\rv) \big] ~, \label{KineticEnergyDensity} \\
    \hat{v}(\rv)&= v(\rv)\hat n(\rv)~,
\label{PotentialEnergyDensity} \\   
    \hat u_0 (\rv) & = \frac{1}{2}\hat \Phi^\dagger(\rv) \ind{3}{s}~ \Phi^\dagger(\sv) w(|\rv - \sv|) \hat\Phi(\sv) \hat\Phi(\rv) ~. \label{InteractionEnergyDensity}
  \end{align}
\end{subequations}
It is easy to verify that $\hat Q_0 = \ind{3}{r}~\hat q_0(\rv)$ and $\hat H_0 = \ind{3}{r}
~\hat h_0(\rv)$.

Before proceeding, we must observe at this point that the thermal energy density (or the energy density, for that matter)  is not unique.  For example, focusing on the kinetic energy density, the expression
\be \label{t1}
\hat{t}_1(\rv)= -\frac{\hbar^2}{4m} \left( \hat \Phi^\dagger(\rv) \big[\nablav^2\hat\Phi(\rv) \big]
+ \big[\nablav^2 \hat \Phi^\dagger(\rv)  \big] \hat\Phi(\rv) \right) ~,
\ee
is \emph{a priori} as good as the expression
\be \label{t2}
\hat{t}_2(\rv)= \frac{\hbar^2}{2m} \big[ \nablav \hat \Phi^\dagger(\rv) \big] \cdot \big[ \nablav \hat \Phi(\rv) \big]~.
\ee
Both expressions integrate to the total kinetic energy $\hat T$, even though they are obviously different at each point.  On the one hand, the $t_1$ form has the advantage that it can be derived from the non-relativistic limit of the time-component of the relativistic stress tensor for Dirac electrons,  the latter being arguably a physical quantity, namely, the source of the gravitational field. On the other hand $t_2$ is strictly positive, which is intuitively expected of a kinetic energy.  More importantly, the $t_2$ form will allow us to construct an energy current that has a simple scaling property~\cite{QinShi:11} upon introduction of the Luttinger potential, discussed below.   In the following we will therefore adopt the $t_2$ form of the kinetic contribution to the thermal energy density.

\subsection{Thermal potential} \label{SEC:localTemperature}

Following an original idea introduced by Luttinger, we will now consider the effect of an external field $\beta\psi(\rv)$ (for the time being, static) that couples to the thermal energy density.
The perturbed thermal Hamiltonian is
\be\label{KLuttinger}
\hat Q = \ind{3}{r} \hat q_0(\rv) [1+ \psi (\rv)]\,.
\ee 
From the form of the thermodynamic potential it should be evident that, as long as equilibrium conditions are maintained, a variation in the Luttinger potential $\psi(\rv)$ corresponds to a local variation of the temperature:
\be\label{Psi-Static}
\delta \psi(\rv) = \frac{\delta \beta}{\beta} \simeq -\frac{\delta T}{T} ~.
\ee
In the next section we will show that, under such equilibrium conditions, the relation between $\psi(\rv)$ and the local energy density is invertible so we can take $\psi (\rv)$ as an indicator of the temperature that would correspond to a given local energy density if the system were in equilibrium. 

However, the real purpose of Luttinger's potential is to drive the system out of equilibrium, simulating the effect of a temperature gradient in driving a thermal current.  To accomplish this we must allow $\psi$ to be time-dependent, i.e., we have $\psi = \psi(\rv,t)$ and the Hamiltonian $\hat Q$ becomes time-dependent.   It turns out that, {\it under non equilibrium conditions}, the time-dependent Luttinger potential acts as a mechanical proxy for a temperature gradient, i.e., $\delta \psi(\rv,t) \simeq \delta T/T$. Notice the change in sign compared to the equilibrium expression~(\ref{Psi-Static}).   The identification of $\delta \psi$ with $\delta T/T$ under non-equilibrium conditions is simple, but surprisingly subtle.
The point is that under equilibrium conditions we have both a non-uniform $\psi$ and a non-uniform temperature, which is associated with the nonuniform thermal energy density $q_0(\rv)$,  and the thermal current is zero.  At the same time, we can think of the equilibrium state as the net result of two non-equilibrium processes, which exactly cancel each other.  More precisely, the thermal current driven by $\delta \psi$  must be the opposite of the thermal current driven by $\delta T/T$, so that the sum of the two currents is zero (notice that this is essentially the argument leading to Einstein's relation between diffusion constant and conductivity).  But we have already seen that $\delta \psi$ is the opposite of $\delta T/T$: it follows that the thermal conductivity, which relates the thermal current to $\delta T/T$, is exactly the same as the thermal conductivity that relates the thermal current to $\delta \psi$.  In this precise sense $\delta \psi$ acts as a mechanical proxy of $\delta T/T$ ~\cite{Shastry:09}.  We will often refer to $\psi$ as to the ``thermomechanical potential" in what follows.

\begin{figure}
  \begin{center}
    \includegraphics[width=.5\textwidth]{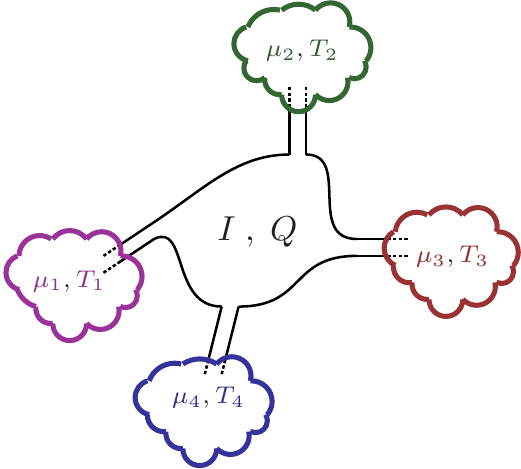}
  \end{center}
  \caption{Sketch of a typical transport setup. Various leads with different chemical potentials and temperatures are connected to a nano scale device. The differences in chemical potentials and temperatures induce a charge , $I$, and heat current, $Q$, through the device.}
  \label{FIG:transportSetup}
\end{figure}
In practical applications we consider an electronic device that is connected by leads and thermal contacts to a set of thermal reservoirs, which are separately in equilibrium at chemical potentials $\mu_i$ and temperatures $T_i$, as shown in Fig.\ \ref{FIG:transportSetup}.  In analogy with the theory of mesoscopic electric transport, we will require some of the reservoirs to have fixed temperature and inject a thermal current (more precisely defined below) into the system, while other reservoirs do not exchange any current and their temperatures are allowed to ``float" to a local equilibrium value.  The first type of reservoir is a model for a thermal source, while the second type is a model for a local temperature  probe~\cite{EngquistAnderson:81}.  The standard treatment of this electro-thermal device is based on the Landauer-B\"uttiker (LB) multi terminal formalism, in which the electric and thermal current are expressed in terms of transmission probabilities of electrons from one reservoir into another.

To treat this system employing Luttinger's thermomechanical potential we have two possibilities~\cite{EichVignale:14b}. The first possibility (method I), which closely mimics the LB approach, is to start with the reservoirs in equilibrium with static thermomechanical fields $\psi_i$.  It is clear that these fields produce in each reservoirs the same populations that would be produced by the temperatures $T_i$ in the LB approach. At the initial time $t=0$ the $\psi$ potentials are turned off in all the reservoirs and the subsequent evolution of the system is tracked.  This evolution takes place under the action of the unperturbed Hamiltonian $\hat Q_0$ ($\psi = 0$)  just as in the LB model, but with initial populations that were dictated by the temperatures $T_i$.  In Ref.\ \cite{EichVignale:14b} we have shown that the long-term or steady-state currents that flow in the reservoirs are identical to those obtained in the LB approach.

The second possibility (method II) is to start with all the reservoirs at the same temperature.  At the initial time $t=0$ different thermomechanical potentials are turned on in the reservoirs, and the subsequent evolution of the system is calculated.  The physical situation is now different from the LB setup, because the initial populations reflect equal temperatures in all reservoirs, but the system evolves under the action of persistent thermomechanical potentials in the leads.  Long-term or steady-state currents are established, in the attempt to equilibrate with the now constant value of $\psi$.  The results are different from those of the LB method, which, as just described, is mimicked by method I. However, in the linear regime both methods are identical provided we identify $\delta \psi_\mathrm{II} = \delta T / T = -\delta \psi_{\mathrm{I}}$. The differences that appear under strong bias arise entirely from the persistent modification of the Hamiltonian via $\psi$. It is precisely in this situation that one must ask the question: does the thermomechanical potential $\psi$ have a genuine physical meaning, or is it just a formal device to calculate thermal responses?  Are there any physical situations in which modeling the system in terms of a persistent $\psi$ potential, i.e., method II, would be more appropriate than method I?

The answers to these questions are not entirely clear at this time.  It seems plausible that the application of a persistent thermomechanical potential would be an appropriate description for an adiabatic heating process, in which the temperature is changed not by heat transfer, i.e., repopulation of energy levels, but by an actual mechanical modification of the energy levels. 

\subsection{Equations of motion} \label{SEC:eom}

Let us now consider more closely the form of the thermal or heat current. This is identified with the entropy current density times the temperature.  One way to proceed is to calculate the time derivative of the entropy density operator, which we take to be
\be
\beta^{-1} \hat s(\rv)= \hat q(\rv)-\omega(\rv)\,.
\ee
The local thermodynamic potential density $\omega(\rv)$ is just a number, which does not contribute to the equation of motion. It follows that
\be \label{heatConservation}
\beta^{-1}\partial_t \hat s(\rv) = \partial_t \hat q(\rv) \equiv -\nablav \cdot \jv_{q}(\rv)\,,
\ee
where the last equality is our definition of the heat current, which is given by the combination of energy current density and particle current density:
\be
\jv_{q}(\rv)\equiv\jv_{h}(\rv)-\mu\jv_{n}(\rv)
\ee

Since we are now in the presence of the thermomechanical potential both the energy and thermodynamic potential density are multiplied by the renormalization factor $1+\psi$ in the following manner
\ber
q(\rv)&\equiv& q_0(\rv)[1+\psi(\rv)]\,,\nonumber\\
\omega(\rv)&\equiv& \omega_0(\rv)[1+\psi(\rv)]\,.
\eer
Also the particle current density is modified in a similar way as can be established by computing the continuity equation in the presence of the coupling to $\psi$. Since only the commutator of the density and the kinetic energy contributes, and the thermomechanical potential enters as a ``mass renormalization'' $m^\star / m = 1 + \psi$ in the kinetic energy, this leads to
\be\label{CurrentDensityPsi}
\hat \jv_{n} (\rv) = [1+\psi(\rv)]~ \hat\jv_{n,0}(\rv) ~,
\ee

The expression for the thermal current density is more convoluted.  Because this current contains both an energy and a velocity, one would hope for it to scale with the local factor $[1+\psi(\rv)]^2$.  This can be achieved for the current of kinetic energy in the $t_2$ form, and for the external potential energy current, but not for the interaction energy density current, which has a more complicated nonlocal structure. 
The complete energy current density, $\jv_h(\rv)$, is given by
\be
\hat\jv_{h}(\rv) =\hat\jv_t(\rv)+\hat\jv_v(\rv)+\hat \jv_u(\rv)+\hat\jv_f(\rv)\,.
\ee
The first term on the right hand side,
\be\label{JTCurrent}
\left[ \jv_{t}(\rv) \right]_i \equiv- [1+\psi(\rv)]^2\frac{i \hbar^3}{8m^2} 
\left( \big[\partial_{i}\hat\Phi^\dagger \big] \big[\nablav^2 \hat \Phi \big] 
- \big[ \nablav^2 \hat \Phi^\dagger \big] \big[ \partial_{i} \hat \Phi \big] -
 \big[\partial_{i}\nablav \hat\Phi^\dagger \big] \cdot \big[ \nablav \hat \Phi \big] 
- \big[\nablav \hat \Phi^\dagger \big] \cdot\big[ \partial_i \nablav \hat \Phi \big] \right) ~.
\ee 
is the kinetic energy contribution. A detailed derivation of this result is presented in Appendix~\ref{APP:eomKin}.  
The potential energy current is given by
\be\label{JVCurrent}
\hat \jv_{v}(\rv)\equiv[1+\psi(\rv)]^2v(\rv)  \hat \jv_{n,0}(\rv)~,
\ee
and is derived in Appendix~\ref{APP:eomPot}.
The remaining terms arise from the electron-electron interaction, and their  expressions, are given in Eqs.~(\ref{JUCurrent}) and ~(\ref{JFCurrent}) of Appendix~\ref{APP:eomInt}.  In particular, $\jv_{u}$ is a ``convective" current of interaction energy carried by a moving volume element of the electron fluid: this current satisfies the local scaling with $(1+\psi)^2$  (see Eq.~(\ref{JUCurrent})).  The last term, $\jv_{f}$, describes the work done on the volume element by the surrounding medium and does not satisfy the local scaling (see Eq.~(\ref{JFCurrent})).  

Let us consider, for orientation, the case of a homogeneous interacting electron gas of uniform density $n_0$, energy density $\varepsilon_0$, and thermodynamic potential density  $\omega_0$.  In equilibrium all currents vanish.  Let us now perform a Galilean transformation to a reference frame with velocity $-\vv$ so that, in the new reference frame, all the electrons are imparted a positive velocity $\vv$.  In the new reference frame it is a simple exercise to show that the currents are
\be
\jv_{n} = n_0 \vv
\ee
and
\be
\jv_{h} = (\varepsilon_0 + p_0)\vv
\ee
where $p_0=-\omega_0$ is the pressure.  Then we can immediately verify that
\be\label{UniformCurrents}
\jv_{q}=\jv_{h}-\mu\jv_{n} =(\varepsilon_0-\mu n_0-\omega_0)\vv  = \beta^{-1} n_0 s_0 \vv\,,
\ee
where $n_0s_0$ is the entropy density (thus, $s_0$ is entropy per particle). This confirms our surmise that the $\jv_q$ current (here calculated in the absence of the thermomechanical potential $\psi$) is indeed the entropy current.

Up to this point we have closely followed Luttinger's approach to mechanically simulate a thermal gradient.  This is a sufficient basis for the time-dependent DFT that we construct in the next section.  Before closing this section, however, we want to briefly comment on a different, and interesting approach, that has recently been proposed by Tatara~\cite{Tatara:15a}, motivated by calculations of the Nernst effect (Hall effect driven by a thermal gradient)~\cite{QinShi:11}.

The basic idea is that the Luttinger interaction term
\be\label{LuttingerInteraction}
\hat Q_\psi = \ind{3}{r}~ \hat q_0(\rv) \psi(\rv,t)
\ee
can be eliminated, to first order in $\psi$,  in favor of an interaction with a thermal vector potential $\Av_{th}(\rv,t)$:
\be\label{KA}
\hat Q_\Av =\ind{3}{r}~\hat \jv_{q}(\rv) \cdot \Av_{th}(\rv,t)\,.
\ee
The relation between the thermal vector potential $\Av_{th}$ and the Luttinger potential $\psi$ is
\be
\partial_t \Av_{th}(\rv,t) = -\nablav \psi(\rv,t)\,.
\ee
The reader will recognize the similarity between this transformation and the familiar gauge transformation of electrodynamics, in which the scalar electric potential $\phi$, coupling to the charge density,  can be eliminated in favor of a vector potential $\Av$, coupling to the electric current density, such that $\partial_t \Av = -\nablav\phi$.
However, the gauge transformation of electrodynamics is exact, while the present transformation is approximate and becomes exact only to linear order in the Luttinger potential.
To see that this is the case, it is sufficient to apply to the Luttinger Hamiltonian of Eq.~(\ref{KLuttinger}) the unitary transformation
\be
\hat U(t) = e^{-i \ind{3}{r}~\hat q_0(\rv) \chi(\rv,t)}\,.
\ee 
where $\chi(\rv,t)\equiv \int_0^t dt'~\psi(\rv,t')$ and $-\nablav \chi(\rv,t)=\Av_{th}(\rv,t)$.  
Under this transformation the hamiltonian becomes
\be
\hat Q' = \hat U^\dagger(t) \hat Q \hat U(t) -i \hat U^\dagger(t) \partial_t \hat U(t)
\ee
The second term of this expression cancels the Luttinger interaction $\hat Q_\psi$ (Eq.~(\ref{LuttingerInteraction})), while the first term, applied to first order in $\psi$, generates the coupling $\hat Q_\Av$ (Eq.~(\ref{KA}))  between the thermal current and the thermal vector potential.  Up to this point, we have simply made a transformation that may present some technical advantages~\cite{Tatara:15a}.  However, we can also treat the thermal vector potential as a driving field in its own right, in which case it will have not only a longitudinal component (equivalent to the Luttinger's thermomechanical potential), but also a transverse component.  The interest of such transverse components is that they give us control on the transverse components of the thermal current: in other words, an interaction of the form~(\ref{KA}) could be used as the basis for a thermal {\it current} density functional theory, in which the {\it full} thermal current, with longitudinal and transverse components, becomes a basic variable.  We will not pursue this possibility in this review.  We will formulate our thermal density functional theory in terms of the scalar thermomechanical potential. The resulting theory determines, in principle exactly, the energy density and the longitudinal part of the thermal current (i.e. the divergence of the thermal current), from which the practically important thermal energy fluxes can be computed.

\section{Structure of thermal DFT}

\subsection{Static thermal Density-Functional Theory} \label{SEC:staticThermalDFT}

Thermal DFT aims at the description of thermoelectric transport phenomena and therefore is a time-dependent or nonequilibrium theory. There are, however, two scenarios in which a static thermal DFT is of interest: 1) For situation in which already the initial state of the system is exposed to nonuniform temperatures. 2) For the derivation of so-called adiabatic approximations. For both it is required to consider a generalized equilibrium theory. We will refer to this theory as quasi-equilibrium theory due to the fact that is time-independent, but allows for a non-vanishing static thermomechanical potential.

\subsubsection{Quasi-equilibrium grand potential} \label{SEC:QuasiEquilibriumGrandPotential}

Let us consider a the quasi-equilibrium grand potential as functional
of the density matrix, $\hat{D}$,
\begin{align}
  \Omega[\hat{D}] = \trace\left\{\hat{D} \left[\hat{Q}_{v,\psi}
  + \frac{1}{\beta} \log\hat{D} \right]\right\} ~, \label{GPF}
\end{align}
with Hamiltonians of the form
\begin{align}
  \hat{Q}_{v,\psi} = \ind{3}{r} \big[1 + \psi(\vr)\big] \hat{q}(\vr)
  = \ind{3}{r} \big[1 + \psi(\vr)\big] \left(\hat{h}(\vr) + (v(\vr) - \mu) \hat{n}(\vr) \right) ~, \label{quasiEquilibriumHamiltonian}
\end{align}
The chemical potential, $\mu$, and the inverse temperature, $\beta=(\kB T)^{-1}$,
define the global zero and scale of the energy, respectively. Mermin~\cite{Mermin:65} showed that
\begin{align}
  \Omega[\hat{D}_{\tilde{v},\psi}] < \Omega[\hat{D}] ~, \label{GPFvariationalPrinciple}
\end{align}
with the quasi-equilibrium grand-canonical-ensemble density matrix given by
\begin{align}
  \hat{D}_{\tilde{v},\psi} = \exp\left(- \beta \hat{H}_{\tilde{v},\psi} \right) / 
  \trace\left\{ \exp\left(- \beta \hat{H}_{\tilde{v},\psi} \right)\right\} ~, \label{Dqeq}
\end{align}
where we have introduced
\begin{subequations}
  \begin{align}
    \hat{H}_{\tilde{v},\psi} & =\ind{3}{r} \Big[ (1 + \psi(\rv)) \hat h (\rv) + \tilde v(\rv) \hat n(\rv) \Big] ~, \label{Hvtpsi} \\
    \tilde{v}(\vr) & = [1 + \psi(\vr)] [v(\vr) - \mu] ~. \label{vt}
  \end{align}
\end{subequations}
It is important to stress that in Mermin's proof of Eq.\ \eqref{GPFvariationalPrinciple} no assumption
about any particular Hamiltonian is required.\footnote{For details we refer to the original proof given in the
appendix of \onlinecite{Mermin:65}.} Accordingly the proof is valid also for the
quasi-equilibrium Hamiltonians defined in Eq.\ \eqref{quasiEquilibriumHamiltonian}.
Using the corresponding quasi-equilibrium density matrix, Eq.\ \eqref{Dqeq},
we can write the grand potential as functional
of the external potential, $\tilde{v}(\vr)$ and the thermomechanical potential
$\psi(\vr)$, i.e.,
\begin{align}
  \tilde{\Omega}[\tilde v,\psi] & = - \tfrac{1}{\beta}
  \log\trace\left\{e^{-\beta\ind{3}{r}\left[(1+\psi(\vr))\hat h(\vr)
  + \tilde v(\vr) \hat n(\vr) \right]} \right\} . \label{GPF_v_psi}
\end{align}
Moreover, we know the density and energy density as functionals of the potentials,
\begin{subequations}
  \begin{align}
    n(\vr) = \trace\Big\{D_{\tilde{v},\psi} \hat{n}(\vr) \Big\}
    ~, \label{n_v_psi} \\
    h(\vr) = \trace\Big\{D_{\tilde{v},\psi} \hat{h}(\vr) \Big\}
    ~. \label{h_v_psi}
  \end{align}
\end{subequations}

\subsubsection{Uniqueness of the static potentials--density mapping} \label{SEC:StaticThermalDFTMapping}

Let us now define the potentials
\begin{subequations}
  \begin{align}
    \tilde{v}_\lambda(\vr) & = (1 - \lambda) \tilde{v}_0(\vr)
    + \lambda \tilde{v}_1(\vr) ~, \label{vLambda} \\
    \psi_\lambda(\vr) & = (1 - \lambda) \psi_0(\vr)
    + \lambda \psi_1(\vr) ~, \label{psiLambda}
  \end{align}
\end{subequations}
and denote the corresponding quasi-equilibrium density matrices
by $\hat{D}_\lambda$. Due to the variational principle
\eqref{GPFvariationalPrinciple} we have
\begin{align}
  & \tilde{\Omega}[(1-\lambda)\tilde{v}_0 + \lambda \tilde{v}_1,
  (1-\lambda) \psi_0 + \lambda \psi_1]
  = \trace\left\{ \hat{D}_\lambda \left[ 
  \hat{H}_{\tilde{v}_\lambda,\psi_\lambda}
  + \frac{1}{\beta} \log\hat{D}_\lambda \right]\right\} \nn
  & = (1 - \lambda) \trace\left\{ \hat{D}_\lambda \left[ 
  \hat{H}_{\tilde{v}_0,\psi_0}
  + \frac{1}{\beta} \log\hat{D}_\lambda \right]\right\}
  + \lambda \trace\left\{ \hat{D}_\lambda \left[ 
  \hat{H}_{\tilde{v}_1,\psi_1}
  + \frac{1}{\beta} \log\hat{D}_\lambda \right]\right\} \nn
  & > (1 - \lambda) \trace\left\{ \hat{D}_0 \left[ 
  \hat{H}_{\tilde{v}_0,\psi_0}
  + \frac{1}{\beta} \log\hat{D}_0 \right]\right\}
  + \lambda \trace\left\{ \hat{D}_1 \left[ 
  \hat{H}_{\tilde{v}_1,\psi_1}
  + \frac{1}{\beta} \log\hat{D}_1 \right]\right\} \nn
  & = (1 - \lambda)\tilde{\Omega}[\tilde{v}_0,\psi_0]
  + \lambda \tilde{\Omega}[\tilde{v}_1, \psi_1] ~, \label{ConcavityProof}
\end{align}
which proves that $\tilde{\Omega}[\tilde{v}, \psi]$ is a strictly concave functional
of the potentials. 
Hence we can employ a Legendre transformation to obtain a
\emph{convex} functional ${F[n,h]}$, i.e., a functional of the
density, $n(\vr)$, and energy density $h(\vr)$, which are the conjugate
variables to the external potentials $\tilde{v}(\vr)$ and $\psi(\vr)$,
\begin{subequations} \label{den_pot}
  \begin{align}
    n(\vr) & = \fder{\tilde{\Omega}[\tilde{v},\psi]}{\tilde{v}(\vr)}
    ~, \label{nr} \\
    h(\vr) & = \fder{\tilde{\Omega}[\tilde{v},\psi]}{\psi(\vr)} ~. \label{hr}
  \end{align}  
\end{subequations}
More precisely: $F[n,h]$, the quasi-equilibrium free energy, is defined as the
negative of the Legendre transform of the grand potential Eq.\ \eqref{GPF_v_psi},
\begin{align}
  F[n,h] & = \tilde{\Omega}[v[n, h],\psi[n, h]]
  - \ind{3}{r} \left( \psi[n,h](\vr) h(\vr) + v[n, h](\vr) n(\vr)\right)
  ~.  \label{F_n_h} 
\end{align}
Equations.\ \eqref{nr} and \eqref{hr} define the densities $n(\vr)$
and $h(\vr)$ as functionals of the external potentials $\tilde{v}(\vr)$
and $\psi(\vr)$. The concavity of $\tilde{\Omega}[\tilde{v},\psi]$
guarantees that we can invert these functional mappings to yield
the external potentials as functionals of the densities, which, in fact,
is the analogue of Mermin's finite temperature DFT (FT-DFT) mapping theorem~\cite{Mermin:65} for static thermal DFT.
From Eq.\ \eqref{F_n_h} follows immediately that
\begin{subequations} \label{pot_den}
  \begin{align}
    \tilde{v}(\vr) & = - \fder{F[n,h]}{n(\vr)} ~, \label{vr} \\
    \psi(\vr) & = - \fder{F[n,h]}{h(\vr)} ~. \label{psir}
  \end{align}  
\end{subequations}

The quasi-equilibrium free energy functional ${F[n,h]}$ of static thermal
DFT is connected to the free energy functional ${F^{\mathrm{eq}}_\beta[n]}$
of FT-DFT by,
\begin{align}
  F^\eq_\beta[n] & = F[n,h^\eq_\beta[n]] ~. \label{Feq_F}
\end{align}
In equation \eqref{Feq_F} we have implicitly defined the equilibrium energy density (at inverse temperature $\beta$)
as functional of the density, $h^\eq_\beta(\vr) = h^\eq_\beta[n](\vr)$, using Mermin's FT-DFT mapping theorem.

\subsubsection{Constrained Search Formulation} \label{SEC:ConstrainedSearch}

We conclude this section by presenting an alternative way to define
the free-energy functional of static thermal DFT. We start
by rewriting the grand potential Eq.\ \eqref{GPF_v_psi} as
\begin{align}
  \tilde{\Omega}[\tilde{v},\psi] = 
  \min_{\hat{D}} \left\langle \ind{3}{r}((1+\psi(\vr))\hat h(\vr)
  +\tilde{v}(\vr)\hat n(\vr))+\tfrac{1}{\beta}\log\hat{D} \right\rangle
  ~, \label{O_v_psi_CS}
\end{align}
where the minimization is taken over all statistical operators $\hat{D}$
normalized to ${1}$. Now we employ the so-called 
\emph{constrained-search formalism}, which is an alternative route
to define functionals in static DFT.\cite{Levy:82,Lieb:83}
The idea of the constrained-search procedure is to split
the minimization process into two steps:
1) Minimize over all statistical operators yielding prescribed densities
$n(\vr)$ and $h(\vr)$.
2) Minimize over all $n(\vr)$ and $h(\vr)$.
This leads to
\begin{align}
  \tilde{\Omega}[\tilde{v},\psi] =
  \min_{n(\vr),h(\vr)} \ind{3}{r} \left( \psi(\vr)h(\vr)
  + \tilde{v}(\vr)n(\vr) \right) + F[n, h]  ~, \label{O_F_LT}
\end{align}
where
\begin{subequations} \label{constrainedSearch}
  \begin{align}
    F[n, h] & = \min_{\hat{D} \to n(\vr),h(\vr)} \left\langle \ind{3}{r} \hat h(\vr)
    + \tfrac{1}{\beta} \log\hat{D} \right\rangle \nn
    & = \ind{3}{r} h(\vr) - \tfrac{1}{\beta} S[n,h] , \label{F_n_h_S_n_h} \\
    S[n,h] & = \max_{\hat{D}\to n(\vr),h(\vr)}
    - \left\langle \log\hat{D} \right\rangle ~. \label{S_n_h} 
  \end{align}
\end{subequations}
Eq.\ \eqref{O_F_LT} emphasizes that ${-F[n, h]}$ and $\Omega[v,\psi]$ are
related by a Legendre transformation and the set of potentials
$[ \tilde{v}(\vr),\psi(\vr) ]$ and the set of densities $[ n(\vr),h(\vr) ]$ form
a pair of conjugated variables. Moreover,
Eq.\ \eqref{F_n_h_S_n_h} highlights that the \emph{universal}
free-energy functional ${F[n, h]}$ consists of a (trivial) part
linear in the energy density and the entropy (times the reference temperature) as a functional of
the densities. The entropy functional is defined as the maximum
entropy compatible with the prescribed densities $n(\vr)$ and $h(\vr)$.
This definition of the entropy should be contrasted with equilibrium FT-DFT,
where the entropy, as functional of the density alone, cannot be defined
in terms of a constrained search, because the energy contribution to the free energy
is no trivially given in terms of the density. Instead, only the free energy can be defined via
a constrained search procedure,
\begin{align}
  F^\eq_\beta[n] = \min_{\hat{D} \to n(\vr)} \left\langle \ind{3}{r} \hat h(\vr)
  + \tfrac{1}{\beta} \log\hat{D} \right\rangle ~. \label{FeqConstrainedSearch}
\end{align}

\subsection{Time-dependent thermal Density-Functional Theory} \label{SEC:ThermalDFT}

Transport phenomena are intrinsically out of equilibrium and therefore a time-dependent description is required~\cite{DiVentra:08}. In the previous section we have explained the formal framework of static thermal DFT. Here, we describe the full-fledged time-dependent thermal DFT, suitable to address transport phenomena including the effects of retardation or history dependence.

\subsubsection{The action functional in thermal DFT} \label{SEC:ActionFunctional}

\begin{figure}
  \begin{center}
    \includegraphics[width=.5\textwidth]{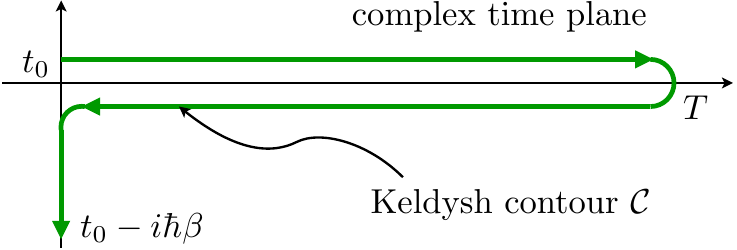}
  \end{center}
  \caption{Integration contour for the Keldysh action of Eq.\ \eqref{AF_v_psi}.}
  \label{FIG:KeldyshContour}
\end{figure}

In order to setup the basic functionals we start by promoting the quasi-equilibrium grand potential, which can be viewed as generating functional for the quasi-equilibrium density and energy density, to an action functional,
\begin{align}
  \tilde{\Lambda}[\tilde{v},\psi] = i \hbar \log\trace\left\{ \mathrm{T}_\tau e^{-\frac{i}{\hbar}
  \intc \ind{3}{r}\left((1+\psi(\vr, \tau))\hat h(\vr)
      + \tilde v(\vr, \tau) \hat n(\vr) \right)} \right\} ~. \label{AF_v_psi}
\end{align}
Formally definitions Eq.\ \eqref{AF_v_psi} and \eqref{GPF_v_psi} are quite similar. The biggest difference is that the potentials $\tilde{v}(\vr, \tau)$ and $\psi(\vr, \tau)$ are time-dependent potentials. Therefore we have an additional time integral $\intc\equiv\intcdef$ in the exponent which runs along the so-called Keldysh contour $\mathcal{C}$, depicted in Fig.\ \ref{FIG:KeldyshContour}, in the complex time plane~\cite{Keldysh:64_original,Keldysh:65,StefanucciVanLeeuwen:13}. In Eq.\ \eqref{AF_v_psi} we have parametrized the contour by a real parameter $\tau$, i.e., $\mathcal{C}=t(\tau)$, and the contour ordering $\mathrm{T}_\tau$ is defined w.r.t.\ the parameter $\tau$. Note that we formally write the potentials as functions of the parameter $\tau$, which allows them to be different for the same physical time depending on whether $\tau$ is on the forward or backward running part of the Keldysh contour.
The class of physical potentials is given by $v(\vr, \tau)$ and $\psi(\vr, \tau)$ that are constant in time along the vertical branch of the contour $\mathcal{C}$ and have an identical time-dependence on the part of the contour running forward and backward on the real time axis. In the following we will adapt the convention that we refer to physical potentials whenever the time argument of the potentials is $t$. Moreover, if we evaluate $\tilde{\Lambda}[\tilde{v},\psi]$ for physical potentials the contribution from the horizontal branch actually cancel and we get
\begin{align}
  \tilde{\Lambda}[\tilde{v},\psi] = -i \hbar \beta \tilde{\Omega}[\tilde{v}_\vert,\psi_\vert] ~, \label{AF_GPF}
\end{align}
which means that the value of the action functional is proportional to the grand potential functional introduced in Sec.\ \eqref{SEC:staticThermalDFT}. The subscript ``$\vert$'' in Eq.\ \eqref{AF_GPF} denotes that the $\tau$-independent potential along the verical axis of the Keldysh contour is plugged into the grand potential functional $\Omega[\tilde{v}, \psi]$. The value of the action functional does not provide any further information.

However, the action functional $\tilde{\Lambda}[\tilde{v},\psi]$ is the generating functional for the time-dependent density and energy density, 
\begin{subequations} \label{den_pot_td}
  \begin{align}
    n(\vr, t) & = \fder{\tilde{\Lambda}[\tilde{v},\psi]}{\tilde{v}(\vr, t)} ~, \label{nrt} \\
    h(\vr, t) & = \fder{\tilde{\Lambda}[\tilde{v},\psi]}{\psi(\vr, t)} ~. \label{hrt}
  \end{align}
\end{subequations}
Accordingly $\tilde{\Lambda}[\tilde{v},\psi]$ serves as tool to compute physically interesting expectation values. Moreover, it has been shown~\cite{vanLeeuwen:98} that writing a generating functional employing the Keldysh contour resolves the apparent causality paradox~\cite{vanLeeuwen:98,Vignale:08}, which plagued initial action functionals for TDDFT. 

Now, under the assumption that Eqs.\ \eqref{den_pot_td} can be inverted, we define another action functional via Legendre transformation, i.e., 
\begin{align}
  A[n, h] \equiv \tilde{\Lambda}\big[ \tilde{v}[n, h], \psi[n, \psi] \big]
  - \intc \ind{3}{r} \Big( \psi[n,h](\vr, \tau) h(\vr, \tau) + \tilde{v}[n, h](\vr, \tau) n(\vr, \tau)\Big) ~. \label{A_n_h} 
\end{align}
Note that in Eq.\ \eqref{A_n_h} all functional dependencies imply a history dependence. This means that the potentials $\tilde{v}(\vr, \tau)$ and $\psi(\vr, \tau)$ depend on the density and energy density at earlier times. From Eq.\ \eqref{A_n_h} it is easy to see that
\begin{subequations} \label{pot_den_td}
  \begin{align}
    \tilde{v}(\vr, t) & = - \fder{A[n,h]}{n(\vr, t)} ~, \label{vrt} \\
    \psi(\vr, t) & = - \fder{A[n,h]}{h(\vr, t)} ~. \label{psirt}
  \end{align}
\end{subequations}
Its is crucial to keep in mind that $A[n,h]$ incorporates the dependence on the evolution of both densities, $n(\vr, t)$ and $h(\vr, t)$, on the Keldysh contour. Put differently: the variation of $A[n,h]$ yields time-dependent potentials, $\tilde{v}(\vr, t)$ and $\psi(\vr, t)$, which depend on the history of both densities. However, the similarity to the free energy functional of static thermal DFT (cf.\ Sec.\ \ref{SEC:staticThermalDFT}) suggests a first approximation: Neglecting the history dependence all together. This so-called \emph{adiabatic} approximation reads
\begin{align}
  A^\adia[n,h] & = \intc F[n(\tau), h(\tau)] = \intc \ind{3}{r} h(\vr, \tau)
  - \frac{1}{\beta} \intc S[n(\tau),h(\tau)] \nn
  & = \intc \ind{3}{r} h(\vr, \tau) - \frac{1}{\beta} \bar{S}^\adia[n,h] ~. \label{A_n_h_adia}
\end{align}
The definition of the adiabatic functional implies that the exact action functional can be written as $A[n,h] = A^\adia[n,h] - \frac{1}{\beta} \bar{S}^\dyn[n,h]$. In addition to the adiabatic action, $A^\adia[n,h]$, the dynamical excess entropy, $\bar{S}^\dyn[n,h]$, contributes to the full action. Note that the excess entropy is formally defined as the difference of the full action functional $A[n,h]$ and the adiabatic approximation $A^\adia[n,h]$. We have opted to refer to the remainder as excess entropy, since the adiabatic approximation contains a trivial dependence on the time-dependent energy density. Accordingly we decompose the action functional into
\begin{align}
  A[n,h] = \intc \ind{3}{r} h(\vr, \tau) 
  - \frac{1}{\beta} \left( \bar{S}^\adia[n,h] + \bar{S}^\dyn[n,h] \right)~. \label{A_decomposition}
\end{align}
Only the last part, the dynamical excess entropy $\bar{S}^\dyn[n,h]$, leads to memory, i.e., a dependence of the potentials on the densities at previous times. It is this part that will introduce retardation and dissipation effect in thermal DFT. In App.\ \ref{APP:ThermalResponseInvertibility} we address the issue of the invertibility of Eqs.\ \eqref{den_pot_td} by explicitly deriving the condition under which the linear response functions can be inverted.

\subsubsection{The Kohn-Sham system} \label{SEC:KohnShamSystem}

One key aspect of DFTs is the so-called Kohn-Sham (KS) system~\cite{KohnSham:65}, a system of fictitious non-interacting electrons, which reproduces the density of the interacting system. The construction of the KS system in thermal DFT exhibits subtle differences compared to the usual KS scheme. Virtually every incarnation of DFT requires the densities in the interacting and non-interacting system to be equal. While this is still true for the electronic density, the energy densities of the interacting and non-interacting system are not identical in thermal DFT.

Let us start by noting that the operators yielding the energy density differ for interacting and non-interacting systems,
\begin{subequations} \label{eop}
  \begin{align}
    \hat{h}(\vr) & = \hat{t}(\vr) + \hat{u}(\vr) ~, \label{h} \\
    \hat{h}_\ks(\vr) & = \hat{t}(\vr) ~. \label{hs}
  \end{align}
\end{subequations} 
On the one hand, the operator for the interacting system, $\hat{h}(\vr)$, contains kinetic and interaction contributions, $\hat{t}(\vr)$ and $\hat{u}(\vr)$, respectively. The operator for the non-interacting system, on the other hand, is only the kinetic energy density. Intuitively it seems rather cumbersome to reproduce the interacting energy density, containing interaction and kinetic contributions, with the non-interacting energy density, which is purely kinetic. Instead, considering the standard formulation of DFT, it is natural to write the interacting energy density as the kinetic energy density of the KS system plus the interaction energy density,
\begin{align}
  h(\rv, t) = h_\ks(\rv, t) + \epsilon_\Hxc(\vr, t) ~. \label{h_hs_eHxc}
\end{align}
The energy density $\epsilon_\Hxc(\vr, t)$ contains the Hartree contribution and a contribution due to exchange-correlation (xc) effects,
\begin{subequations} \label{interactionEnergyDensity}
  \begin{align}
    \epsilon_\Hxc(\rv) & = \epsilon_\Hartree(\vr) + \epsilon_\xc(\rv) ~, \label{eHxc_eH_exc} \\ 
    \epsilon_\Hartree(\vr) & = \frac{1}{2} n(\vr,t) v_\Hartree(\vr, t) ~,  \label{eH} \\
    v_\Hartree(\vr, t) & = \ind{3}{r'} n(\vr, t) w(|\rv - \rv') ~. \label{vH}
  \end{align}
\end{subequations}
The formal definition of the interaction energy density employs the energy density as functional of the density for a given reference temperature, which we already introduced at the end of Sec.\ \ref{SEC:StaticThermalDFTMapping}, i.e.,
\begin{align}
  \epsilon_{\Hxc,\beta}[n(t)](\rv) = h^\eq_\beta[n(t)](\vr) - h^\eq_{\ks,\beta}[n(t)](\vr) ~. \label{eHxc}
\end{align}
We can rearrange Eq.\ \eqref{h_hs_eHxc} using definition \eqref{eHxc} in order to introduce the excess energy density,
\begin{align}
  \bar{h}(\vr, t) = h(\vr, t) - h^\eq_\beta[n(t)](\vr) = h_{\ks}(\vr, t) - h^\eq_{\ks,\beta}[n(t)](\vr) ~. \label{hbar}
\end{align}
Equation \eqref{hbar} implies that the connection between the interacting and the non-interacting energy density, given in Eq.\ \eqref{h_hs_eHxc}, requires the excess energy densities to be identical in both systems. Physically this means that the KS system is chosen to be out--of--equilibrium by the same amount as the interacting system. The metric for being out--of--equilibrium is the excess energy density, i.e., the difference of the time-dependent energy density and the instantaneous equilibrium energy density. The instantaneous equilibrium energy density, in turn, is determined by the density common to the interacting and the KS system.

The action functional for the non-interacting KS system, $A_\ks[n,h_\ks]$ can be decomposed analogously to the interacting functional, cf.\ Eq.\ \eqref{A_decomposition}. The difference between the action functional of the interacting and the KS system is the so-called
Hartree-exchange-correlation action,
\begin{align}
  A_\Hxc & = A - A_\ks = \intc \ind{3}{r} \epsilon_{\Hxc,\beta}[n(t)](\vr)
  - \frac{1}{\beta} \left(\bar{S}^\adia_\xc + \bar{S}^\dyn_\xc \right) \nn
  & = E_{\Hxc,\beta}[n](\vr)
    - \frac{1}{\beta} \left(\bar{S}^\adia_\xc + \bar{S}^\dyn_\xc \right)~. \label{AxcDecomp}
\end{align}
Note that we have not specified whether we choose $h$ or $h_\ks$ as our energy-density variable. However, in DFT one usually obtains the expectation values directly from the KS system, which strongly suggests to use the KS energy density, $ h_\ks(\vr, t)$. This means that the xc entropy, i.e., the sum of the adiabatic xc entropy and the dynamic xc entropy, is defined by
\begin{align}
  \bar{S}_\xc[n,h_\ks] = \bar{S}\big[n,h_\ks + \epsilon_\Hxc[n] \big]
  - \bar{S}_\ks\big[ n, h_\ks \big] ~. \label{Sxc_n_hs}
\end{align}
Now we can write out explicitly the functional derivative w.r.t.\ $h_\ks(\vr, t)$ for the xc action.
From Eq.\ \eqref{psirt} it follows immediately that
\begin{align}
  - \frac{1}{\beta} \fder{\bar{S}_\xc[n,h_\ks]}{h_\ks(\vr, t)}
  = \psi_\ks(\vr, t) - \psi(\vr, t) = \bar{\psi}_\xc(\vr, t) ~, \label{psixc}
\end{align}
where the functional derivative is taken at constant density $n(\vr, t)$.

Some care has to be taken when differentiating w.r.t.\ the electronic density at constant $h(\vr, t)$. In definition \eqref{Sxc_n_hs} the energy-density argument of $\bar{S}[n,h_\ks + \epsilon_\Hxc[n]]$ is shifted by an density-dependent amount. From Eq.\ \eqref{vrt} we obtain
\begin{align}
  \tilde{v}_\ks(\vr, t) - \tilde{v}(\vr, t) = \fder{A_\Hxc[n,h_\ks]}{n(\vr, t)} - \ind{3}{r'} \fder{A[n,h_\ks + \epsilon_\Hxc[n]]}{h_\ks(\vrp, t)} \fder{\epsilon_\Hxc[n(t)](\vrp)}{n(\vr, t)} ~. \label{vs_v__AHxc}
\end{align}
The second term on the right hand side is a counter term required to shift the energy argument of the interacting functional inside the density derivative. We point out that this term only appears if we construct explicit approximations for $A_\Hxc[n,h_\ks]$. As we will see later, in the adiabatic local density approximation the potentials are directly approximated and therefore the counter term will not appear. In the present discussion we formally derive the xc potentials from $A_\Hxc[n,h_\ks]$ and, hence, we will keep the counter term. Taking into account decomposition \eqref{AxcDecomp} we arrive at
\begin{align}
  \tilde{v}_\ks(\vr, t) - \tilde{v}(\vr, t) & = \fder{E_\Hxc[n]}{n(\vr, t)}
  - \frac{1}{\beta}\fder{\bar{S}^\adia_\xc[n,h_\ks]}{n(\vr, t)}
  - \frac{1}{\beta}\fder{\bar{S}^\dyn_\xc[n,h_\ks]}{n(\vr, t)} \nn
  & \phantom{=} + \ind{3}{r'} \psi(\vrp, t) \fder{\epsilon_\Hxc[n(t)](\vrp)}{n(\vr, t)} 
  ~. \label{vs_decomposition}
\end{align}
Furthermore, using Eq.\ \eqref{AxcDecomp} to decompose the xc thermomechanical potential, $\bar{\psi}_\xc$, we finally arrive at the expression for the KS potentials:
\begin{subequations}
  \begin{align}
    \tilde{v}_\ks(\vr, t)
    & = \tilde{v}(\vr, t) + v^\eq_\Hxc(\vr, t) + \bar{v}^\adia_\xc(\vr, t) + \bar{v}^\dyn_\xc(\vr, t)
    + \ind{3}{r'} \psi(\vrp, t) \fder{\epsilon_\Hxc[n(t)](\vrp)}{n(\vr, t)}~, \label{vxcDecomp} \\
    \psi_\ks(\vr, t)
    & = \psi(\vr, t) + \bar{\psi}^\adia_\xc(\vr, t) + \bar{\psi}^\dyn_\xc(\vr, t) ~. \label{psixcDecomp}
  \end{align}
\end{subequations}
Both the external potential, $\tilde{v}_\ks(\vr, t)$, and the thermomechanical field, $\psi_\ks(\vr, t)$, have contributions from the adiabatic excess entropy, $S^\adia_\xc[n,h_\ks]$, and the dynamical excess entropy, $S^\dyn_\xc[n,h_\ks]$. The external potential has an additional contribution corresponding to the instantaneous equilibrium potential, 
\begin{align}
  v^\eq_\Hxc(\vr, t) & = \fder{E^\eq_\Hxc[n]}{n(\vr, t)} ~. \label{veq}
\end{align}
The additional contribution due to the density-dependent shift of the energy argument is the potential energy associated with the thermomechanical potential coupled to the change of the instantaneous equilibrium energy density. Also this contribution is an adiabatic contribution which is only determined by the instantaneous density. Therefore we combine this contribution with Eq.\ \eqref{veq} to define
\begin{align}
  \tilde{v}^\eq_\Hxc(\vr, t) = v_\Hartree(\vr, t) + v^\eq_\xc(\vr, t)
  + \ind{3}{r'} \psi(\vrp, t) \fder{\epsilon_\Hxc[n(t)](\vrp)}{n(\vr, t)} ~, \label{vteq}
\end{align}
where we also explicitly extracted the Hartree potential for clarity. This allows us to write the time-dependent KS equation in the compact form:
\begin{align}
  i \hbar \partial_t  \phi_\alpha (\vr,t) =  \left[-\hbar^2 \nablav_{\vr} \frac{1 + \psi(\vr,t) + \bar \psi_\xc(\vr,t)}{2m}  \nablav_{\vr}
  + \tilde v(\vr,t) + \tilde v^\eq_\Hxc(\vr,t) + \bar v_\xc(\vr,t) \right]\phi_\alpha(\vr,t) ~. \label{tdSE}
\end{align}
In Eq.\ \eqref{tdSE} we further combined the adiabatic and dynamical excess contributions of the potential and thermomechanical field for brevity,
\begin{subequations} \label{barPotentials}
  \begin{align}
    \bar{v}_\xc(\vr, t) & = \bar{v}^\adia_\xc(\vr, t) + \bar{v}^\dyn_\xc(\vr, t) ~, \label{vbarDecomp} \\
    \bar{\psi}_\xc(\vr, t) & = \bar{\psi}^\adia_\xc(\vr, t) + \bar{\psi}^\dyn_\xc(\vr, t) ~. \label{psibarDecomp}
  \end{align}
\end{subequations}
The self-consistent solution of the time-dependent Schr\"odinger equation \eqref{tdSE} together with the initial occupations of the orbitals $\phi_\alpha(\vr, t)$ yield the time-evolution of the density $n(\vr, t)$ and the KS energy density, $h_\ks(\vr, t)$. Note that the presence of a thermomechanical potential bears close resemblance with a position dependent mass, which appear, e.g., in the description of compositionally grated semi-conductors~\cite{GellerKohn:93}.

We conclude the section by recapitulating the required approximations for the implementation of the KS scheme in thermal DFT: First of all an approximation of $\epsilon_\Hxc[n](\vr)$ is needed in order to compute the interacting energy density from the KS system. Furthermore $\epsilon_\Hxc[n](\vr)$ implies an approximation for $E^\eq_\Hxc[n]$. Note that $\epsilon_\Hxc[n](\vr)$ is, in principle, a functional already defined in Mermin's FT-DFT. The really new features in the KS equation of thermal DFT are the contributions $\bar{v}_\xc(\vr, t)$ and $\bar{\psi}_\xc(\vr, t)$, which can be obtained from an approximation to the excess entropy. They contain all the retardation effects induced by the interactions. 

\section{Alternative formulations} \label{SEC:AlternativFormulations}

In the previous sections we have treated the phenomenon of heat and energy currents by considering the local temperature as an ``internal'' parameter of the theory. In other words, although the presence of a temperature somehow requires tracing out degrees of freedom of a bath, or a set of baths, we do not consider such a problem as an open one~\footnote{Note that, due to the presence of a temperature, we {\it cannot} classify it as a closed quantum problem either.}. However, alternative approaches have been proposed in the literature that aim at treating explicitly thermoelectricity as an open quantum system problem. Here, we briefly review two of them. One approach solves the stochastic Schr{\"o}dinger equation for the state vector (or, an equation of motion for the density matrix) and reformulates a TDCDFT in such a context. The other employs the stochastic Sch{\"o}dinger equation to describe the coupling of black bodies, i.e., idealized sources of incoherent light, to a quantum system. When the system is coupled to multiple black bodies at different temperatures they induce a thermal gradient and therefore a heat flow in the system. 

\subsection{Stochastic density functional theory for open systems} \label{SEC:StochasticDFT}

There are two ways in which one could formulate a TDCDFT for open quantum systems: via an equation of motion for the density matrix~\cite{BurkeGebauer:05} or an equation of motion for the state vector~\cite{DagostaDiVentra:07}. Since the Hamiltonian of DFT is dependent on the density and/or current density, it depends on the state of the system, and hence it is generally stochastic~\cite{DagostaDiVentra:13}. In this case then, a closed form of the equation of motion for the density matrix is not available, and one is forced to start with the equation of motion for the state vector~\cite{DagostaDiVentra:07}. In the absence of memory, such an equation takes the form ($\hbar=1$)
\begin{equation}
d|\Psi\rangle=\left[-i\hat H |\Psi\rangle -\frac12 \hat V^\dagger
\hat V |\Psi\rangle\right]dt + \hat V |\Psi \rangle dW,
\label{stochasticse-fin}
\end{equation}
where $dW$ is a infinitesimal Wiener process (It\^o calculus is employed), and $\hat V$ is an operator that describes the interaction of the system with an environment (generalization to many environments simply adds a sum over different Wiener processes in Eq.~(\ref{stochasticse-fin})). The Hamiltonian $\hat H$ is a general many-body Hamiltonian in the presence of an external vector potential 
$A(\bf{r},t)$.

The theorem of stochastic TDCDFT~\cite{DagostaDiVentra:07} then states that there is a one-to-one correspondence between the ensemble-averaged current density $\langle \overline{\hat j(r,t)}\rangle$ and the external vector potential, thus leading to the set of Kohn-Sham equations 
\begin{equation}
d|\Psi_{KS}\rangle =\left(-i\hat H_{KS}-\frac12 \hat V^\dagger \hat
V\right)|\Psi_{KS}\rangle dt+\hat V |\Psi_{KS}\rangle dW
\label{ksse}
\end{equation}
where $|\Psi_{KS}\rangle$ is a Slater determinant of single-particle wave-functions and
\begin{equation}
\hat H_{KS}=\sum_{i=1}^N \frac{\left[\hat p_i+eA_\eff(r_i,t)\right]^2}{2m} \label{ksh}
\end{equation}
is the Hamiltonian of non-interacting particles, with $A_\eff(r_i,t)$ an effective exchange-correlation vector potential that is history dependent.

In the absence of interactions (except statistics) Eq.\ \eqref{ksse} is equivalent to the Lindblad equation for the same type of environment~\cite{Lindblad:76}. Note, however, that even for non-interacting particles, due to the presence of a bath (hence of statistical correlations), Eq.\ \eqref{ksse} cannot be reduced to a set of single-particle equations. In Ref.~\cite{PershinDiVentra:08} a single-particle scheme has been proposed that provides particularly accurate results for expectation values of operators that can be written as sums of single-particle operators (such as the density and the current density). Such an approach has been used to compute the thermopower in atomic wires and predict the phenomenon of local temperature oscillations in the absence of interactions~\cite{DubiDiVentra:09}. 

\subsection{Time-dependent thermal transport theory} \label{SEC:TDthermalTransport}

In thermal DFT a proxy for the local temperature is defined by the thermomechanical potential. Loosely speaking this mechanical field controls the local energy of the system under investigation. Therefore, it is meant to represent the exchange of heat or energy with an environment. An alternative approach is to model the environment explicitly. A recent proposal~\cite{BieleRubio:15} accomplishes this by means of (classical) black bodies. According to Planck's law, the radiation emitted by a black body depends only on its temperature. Therefore, they can be used as idealized sources of heat by coupling the black body radiation \emph{locally} to a quantum system. In Ref.\ \cite{BieleRubio:15} the black bodies are treated classically, i.e., it is assumed that a large number of incoherent photons heat up the nanoscale device and there is no back-reaction of the black bodies to the dynamic of the quantum system. In order to allow the quantum system to relax and dissipate energy the entire quantum system is furthermore embedded into a bosonic bath which is held at a constant temperature. 

The quantum system coupled to the black bodies and the bosonic bath can be efficiently simulated via the stochastic Schr\"odinger equation introduced in the previous section. While explaining the details of the implementation of the above setup for thermal transport is beyond the scope of this review, we will highlight two central aspects: 1) Since the black body radiation is treated classically it is straightforward to go beyond the linear response limit by propagating the Schr\"odinger equation for the system explicitly. This means that within this approach one can investigate not only the effect of large temperature gradients, but also large heat currents. In fact, the classical approximation of the photon field of the black bodies implies that we are considering a high photon flux. 2) The bosonic bath allows the system the relax or dissipate energy, such that it reaches a stationary or steady-state, even if the system itself only has a finite number of degrees of freedom.

If the nanoscale device is composed of interacting electrons, the interaction can be taken into account via stochastic TDCDFT. However, within a DFT based on the charge current and density alone the energy transport within the system cannot be addressed directly. In order to describe the charge and energy flow on the same footing, a ``stochastic thermal DFT'' could be envisaged. The combination of using black bodies as physical source for heat with Luttinger's thermomechanical potential may also help to provide an ``ab-initio'' justification for the thermomechanical potential.

\section{Construction of functionals for thermal DFT}

This section is devoted to proposing approximations to the functionals of thermal DFT required for an actual implementation. The strategy for the construction of functionals generalizes the so-called local-density approximation (LDA), which has been proposed already in the seminal paper by Hohenberg and Kohn ~\cite{HohenbergKohn:64}. Furthermore we propose dynamical approximations along the lines of the Vignale-Kohn functional introduced in the context TDCDFT ~\cite{VignaleKohn:96a,VignaleKohn:98}.

\subsection{Adiabatic local density approximation for thermal DFT}

In order to generalize the LDA for thermal DFT we build upon recent progress in determining the free energy of the uniform electron gas. Brown et al.\ employed Quantum Monte Carlo methods to obtain the exchange correlation energy of the uniform electron gas at various densities and temperatures ~\cite{BrownCeperley:13a,BrownCeperley:13b}. Shortly afterwards, based on the aforementioned Quantum Monte Carlo calculations, the parametrization of the free energy of the uniform electron gas as function of the density and the temperature has been refined ~\cite{KarasievTrickey:14}. It turns out that this parametrization not only provides an important step forward in the construction of functionals for Mermin's FT-DFT, but also is all we need for an implementation of the adiabatic LDA (ALDA) in thermal DFT.

In Ref.\ \onlinecite{KarasievTrickey:14} the xc free energy per particle as function of density and temperature. For our purpose we consider the free-energy density as function of density, $n$, and inverse temperature, $\beta$, i.e., $f_\xc(n,\beta)$. Using thermodynamic identities we also have the xc energy density at our disposal,
\begin{align}
  \epsilon_\xc(n,\beta) = f_\xc(n,\beta) + \beta \partial_\beta f_\xc(n,\beta) ~. \label{exc_fxc}
\end{align}
Using $e_\xc(n,\beta)$ we can determine the energy density of the interacting system
in the LDA, i.e.,
\begin{align}
  h = h_\ks + \epsilon_\Hartree[n] + \epsilon_\xc(n,\beta_0) ~, \label{h_hs_lda}
\end{align}
where $h_\ks$ is the local energy density of the KS system (kinetic energy density), $\epsilon_\Hartree[n]$ the Hartree contribution to the local interaction-energy density, and $\beta_0$ the fixed reference temperature. We will see shortly that in the ALDA for thermal DFT we approximate the derivatives of the adiabatic excess entropy independently for the interacting and the non-interacting system. Therefore the counter term introduced in Sec.\ \ref{SEC:KohnShamSystem}, Eq.\ \eqref{vs_v__AHxc} is not present. This means that the ``equilibrium'' contribution to the potential is simply
\begin{align}
  \tilde{v}^\eq_\Hxc \approx v_\Hartree + \partial_n \epsilon_\xc(n,\beta_0) ~. \label{vHxcALDA}
\end{align}

Now we turn to the contribution to the potentials due to the adiabatic excess entropy. In the ALDA they are given by
\begin{subequations} \label{ALDAbarPotentials1}
  \begin{align}
    \bar{v}^\adia_\xc & \approx -\frac{1}{\beta_0} \partial_n s_\xc(n, h_\ks) \\
    \bar{\psi}^\adia_\xc & \approx -\frac{1}{\beta_0} \partial_{h_\ks} s_\xc(n, h_\ks) ~, \label{psi_qeq_LDA}
  \end{align}
\end{subequations}
where the xc entropy density is
\begin{align}
  s_\xc(n, h_\ks) = s(n, h_\ks + \epsilon_\Hxc[n,\beta_0]) - s_\ks(n, h_\ks) ~. \label{sxc_LDA}
\end{align}
Even for the non-interacting electron gas we do not have an explicit expression $s_\ks(n,h_\ks)$. However, we are only interested in obtaining 
\begin{subequations} \label{bmKS}
  \begin{align}
    -\beta_\ks \mu_\ks & = \partial_n s_\ks(n, h_\ks) ~, \label{bsms} \\
    \beta_\ks & = \partial_{h_\ks} s_\ks(n, h_\ks) ~. \label{bs}
  \end{align}
\end{subequations}
Similarly we will evaluate for the interacting electron gas
\begin{subequations} \label{bmI}
  \begin{align}
    -\beta \mu & = \partial_n s(n, h) ~, \label{bm} \\
    \beta & = \partial_{h_\ks} s(n, h) ~, \label{b}
  \end{align}
\end{subequations}
directly. Using Eqs.\ \eqref{bmKS} and \eqref{bmI} we have 
\begin{subequations} \label{ALDAbarPotentials2}
  \begin{align}
    \bar{v}^\adia_\xc & = \frac{\beta\mu - \beta_\ks \mu_\ks}{\beta_0} ~, \label{vbALDA} \\
    \bar{\psi}^\adia_\xc & = \frac{\beta_\ks - \beta}{\beta_0} ~. \label{psibALDA}
  \end{align}
\end{subequations}

First we discuss how Eq.\ \eqref{bmKS} can be solved for the non-interacting electron gas. We know the density and energy density explicitly as function of the chemical potential and the inverse temperature,
\begin{subequations}
  \begin{align}
    n_\ks(\mu,\beta) & = \Big( \tfrac{2 m}{\pi \hbar^2 \beta} \Big)^{3/2}
    \frac{1}{4} \mathrm{I}_{1/2}(\beta \mu) ~, \label{ns_mb} \\
    h_\ks(\mu, \beta) & = \Big(\tfrac{2 m}{\pi \hbar^2 \beta} \Big)^{3/2}
    \frac{3}{8 \beta} \mathrm{I}_{3/2}(\beta \mu) ~, \label{hs_mb}
  \end{align}
\end{subequations}
where we have introduced the complete Fermi integrals
\begin{align}
  \mathrm{I}_{\nu}(t) & = \tfrac{1}{\Gamma(\nu+1)} \indll{x}{0}{\infty} \frac{x^{\nu}}{\exp(x - t) + 1} \sim
  \begin{cases} 
    \tfrac{t^{\nu+1}}{\Gamma(\nu+2)} \;\;\; & \mathrm{for} \;\;\; t \to \infty \\
    \exp(t) \;\;\; & \mathrm{for} \;\;\; t \to -\infty 
  \end{cases} ~, \label{completeFermiIntegrals}
\end{align}
we can find $\mu_\ks$ and $\beta_\ks$ by solving the following equations
numerically:
\begin{subequations}
  \begin{align}
    n & = n_\ks(\mu_\ks,\beta_\ks) ~, \label{Nsolve_n} \\
    h_\ks & = h_\ks(\mu_\ks,\beta_\ks) ~, \label{Nsolve_hs}
  \end{align}
\end{subequations}
where $n$ and $h_\ks$ are the density and energy density of the KS system.

Now we turn to Eq.\ \eqref{bmI}. Also for the interacting electron gas we do not need the entropy density as function of $n$ and $h$, but instead we are only interested in the potential, $\mu$, and inverse temperature, $\beta$, associated with the local density and energy density. In contrast to the non-interacting electron gas we do not have the energy density as function of $\mu$ and $\beta$ at hand, but instead we know the free energy as function of the density and inverse temperature, i.e.,
\begin{align}
  f(n,\beta) = f_\ks(n,\beta) + \epsilon_\Hartree[n] + f_\xc(n,\beta) ~. \label{f_unif}
\end{align}
We can find the inverse temperature by solving the standard thermodynamic relation
\begin{align}
  h & = f(n, \beta) + \beta \partial_\beta f(n, \beta) ~, \label{h_fdf}
\end{align}
for $\beta$, where the energy density on the left hand side has been determined in Eq.\ \eqref{h_hs_lda}. This leads to
\begin{align}
  h_\ks + \epsilon_\Hartree[n] + \epsilon_\xc(n, \beta_0) & =
  h_\ks(\mu_0(n, \beta), \beta) + \epsilon_\Hartree[n] + \epsilon_\xc(n, \beta) \nn
  h_\ks + \epsilon_\xc(n, \beta_0) & =
  h_\ks(\mu_0(n, \beta), \beta) + \epsilon_\xc(n, \beta) ~. \label{Nsolve_b}
\end{align}
Let us carefully analyze Eq.\ \eqref{Nsolve_b}: First of all we see that the Hartree energy is irrelevant for the solution. Secondly, $h_\ks(\mu_0(n,\beta),\beta)$ is the equilibrium energy density of a non-interacting electron gas for a given density $n$ and inverse temperature $\beta$. In Eq.\ \eqref{Nsolve_b} we use that we know the energy density of the non-interacting electron gas explicitly as function of the chemical potential and inverse temperature [cf.\ Eq.\ \eqref{hs_mb}]. However, the chemical potential $\mu_0$, which leads to the density $n$ at inverse temperature $\beta$, has to be determined by solving $n = n_\ks(\mu_0, \beta)$ [cf.\ Eq.\ \eqref{ns_mb}]. Having determined the inverse temperature, $\beta$, of the interacting electron gas we obtain the chemical potential $\mu$ simply by
\begin{align}
  \mu & = \partial_n f(n, \beta) \nn
  & = \partial_n \Big[ f_\ks(n, \beta) + \epsilon_\Hartree[n] + f_\xc(n, \beta) \Big] \nn
  & = \mu_0 + v_\Hartree + \partial_n f_\xc(n, \beta) ~. \label{mu}
\end{align}
Accordingly we arrive at our final form of the adiabatic potentials
\begin{subequations} \label{ALDA}
  \begin{align}
    \tilde{v}^\eq_\Hxc + \bar{v}^\adia_\xc \approx v^\ALDA_\Hxc
    & = \big[ v_\Hartree + \partial_n \epsilon_\xc(n,\beta_0) \big]
    + \frac{\beta}{\beta_0}\big[ \mu_0 + v_\Hartree + \partial_n f_\xc(n, \beta) \big]
    - \frac{\beta_\ks}{\beta_0} \mu_\ks ~, \label{vALDA} \\
    \bar{\psi}^\adia_\xc \approx \psi^\ALDA_\xc & = \frac{\beta_\ks - \beta}{\beta_0} ~, \label{psiALDA}
  \end{align}
\end{subequations}
where $\mu_\ks$, $\beta_\ks$, and $\mu_0$, $\beta$ are obtained by solving
\begin{subequations} \label{Nsolve}
  \begin{align}
    n & = n_\ks(\mu_\ks,\beta_\ks) ~,
    & h_\ks & = h_\ks(\mu_\ks,\beta_\ks) ~, \label{Nsolvemsbs} \\
    n & = n_\ks(\mu_0,\beta) ~,
    & h_\ks + \epsilon_\xc(n, \beta_0) & = h_\ks(\mu_0, \beta) + \epsilon_\xc(n, \beta) ~. \label{Nsolvem0b}
  \end{align}
\end{subequations}
We point out that the implementation of the ALDA in thermal DFT is quite different from the usual implementations of the ALDA in TDDFT. In TDDFT one can employ an explicit parametrization of the xc energy in terms of the density. In thermal DFT the ALDA is based on the LDA for FT-DFT. In fact, the entropy as function of the density and energy density is the Legendre transform of the free energy as function of density and temperature. Instead of providing an explicit parametrization of the entropy, we propose to implicitly determine the potentials via the algorithm described in Eqs.\ \eqref{ALDA} and \eqref{Nsolve}. We conclude by pointing out a caveat in implementing the ALDA in practice: Formally the thermomechanical potential vanishes if the system is in equilibrium. In practice this requires the entropy functional to ``detect'' that the density and the energy density are equilibrium densities. However, this can not be decided locally -- except, of course,  for the uniform gas. This means that the ALDA will yield a nonvanishing
$\bar{\psi}_\xc$ if the system is in a nonuniform equilibrium state. This would lead to spurious dynamics if the initial state is obtained via the LDA of Mermin's FT-DFT. Hence, the initial state for thermal DFT has to be computed within ALDA of thermal DFT. Moreover, it remains to be seen whether the proposed scheme for the ALDA can be carried out efficiently in practice. An important issue to analyze is whether Eqs.\ \eqref{Nsolve} always afford a solution for the density and energy density of an inhomogeneous system.

\subsection{Leading dissipative corrections}

The first step in going beyond the adiabatic approximation is to include the dependence of the effective potentials on the time derivatives of the densities.  Since these time derivatives are related by continuity equations to the divergences of the particle and thermal energy currents, it is natural to express the leading corrections beyond ALDA in terms of current densities.  In the spirit of the local density approximation we adopt the homogeneous electron gas of density $n$ as a reference system and use the thermoelectric conductivity matrix $\tilde{\mat{L}}$ of this system to relate the gradients of the dynamical exchange-correlation potentials, i.e., the xc electric field and the xc thermal gradient field, to the particle and thermal current densities.  We start from the standard linear response formula
\be \label{jejq_ET}
  \begin{pmatrix} \jv_e\\[1em] \jv_q \end{pmatrix} = \begin{pmatrix} \tilde{L}_{11} & \tilde{L}_{12}\\[1em] \tilde{L}_{21} & \tilde{L}_{22} \end{pmatrix} \begin{pmatrix} \Ev \\[1em] -\frac{\nablav T}{T} \end{pmatrix} ~,
\ee
where $\Ev = -\nablav \phi$ is the electric field and $\jv_e = -e \jv_n$ the electric current. The generalized conductivities $\tilde{L}_{ij}$ are determined as follows.  If the temperature is uniform ($\nablav T=0$) then the current is driven exclusively by the electric field via the electric conductivity $\sigma$: this gives $\tilde{L}_{11}=\sigma$.  At the same time, there is a thermal energy current riding on top of the electric current, which is given by $\jv_q = \Pi \jv_e$, where $\Pi$ is the Peltier coefficient, which, in the case of a homogeneous electron gas, equals $-\kB T s_0 / e$ (cf.\ Eq.\ \eqref{UniformCurrents} in Sec.\ \ref{SEC:TMpotential}), where  $s_0$ is the entropy per particle of the electron gas. We thus have $\tilde{L}_{21} = \Pi\sigma$.  The coefficient $\tilde{L}_{21}$ is now determined by Onsager reciprocity~\cite{Onsager:31a,Onsager:31b}, which requires $\tilde{L}_{21} = \tilde{L}_{12}$.  Lastly, consider a thermal conductivity measurement, in which a gradient of temperature is applied and the electric current is zero.  The condition $\jv_e=0$ implies that an electric field  is present, such that
\be
\Ev = \frac{\tilde{L}_{12}}{\tilde{L}_{11}} \frac{\nablav T}{T} = \Pi \frac{\nablav T}{T}
\ee
Then we have the thermal current density
\be
\jv_q = -\left(\tilde{L}_{22} - \sigma \Pi^2 \right) \left(\frac{\nablav T}{T}\right)\,.
\ee
Equating the coefficient of $-\nablav T$ to the thermal conductivity $\kappa$ we obtain
\be
\tilde{L}_{22} =\kappa T + \sigma\Pi^2 ~.
\ee
Thus, the generalized conductivity matrix takes the form
\be
\tilde{\mat{L}} = \begin{pmatrix}
  \sigma \quad & \Pi \sigma \\[1em]
  \Pi \sigma \quad & T \kappa + \Pi^2 \sigma 
\end{pmatrix} ~.
\ee
In the context of Thermal DFT it is useful to rewrite Eq.\ \eqref{jejq_ET} in terms of the \emph{particle} current and the external potential and the thermomechanical potential, i.e.,
\begin{align}
  \begin{pmatrix} \jv_n \\ \jv_q \end{pmatrix}
  & = - \mat{L} \begin{pmatrix} \nablav v \\ \nablav \psi \end{pmatrix} ~, \label{jnjq_VT} \\
  \mat{L} & = \begin{pmatrix} \frac{1}{e^2} \sigma & -\frac{1}{e} \sigma \Pi \\[1em] -\frac{1}{e} \sigma \Pi & T \kappa + \sigma \Pi^2 \end{pmatrix} ~, \label{Ldef}
\end{align}
where we used $\Ev = \frac{1}{e} \nablav v$ and $\frac{1}{T} \nablav T= \nablav \psi$. Its inverse is the resistivity matrix
\be
\mat{L}^{-1} = \begin{pmatrix} \frac{e^2}{\sigma} + \frac{e^2 \Pi^2}{\kappa T} & \frac{e \Pi}{\kappa T} \\[1em]
\frac{e \Pi}{\kappa T} & \frac{1}{\kappa T} \end{pmatrix} ~. 
\ee
The resistivity matrix $\Lv^{-1}$ gives the \emph{dynamical} fields $\nablav v$ and $\nablav\psi$, which are associated with the currents $\jv_n$ and $\jv_q$ according to the formula
\be
  \begin{pmatrix} \nablav v \\ \nablav \psi \end{pmatrix}_\dyn = - \Lv^{-1}  \begin{pmatrix} \jv_n \\ \jv_q \end{pmatrix} ~.
\ee
Electron-electron interactions are fully accounted for in the matrix elements of $\Lv^{-1}$.  We can construct the corresponding matrix $\Lv_\ks^{-1}$ for the {\it non-interacting} electron gas at the same homogeneous density.  This gives
\be
  \begin{pmatrix} \nablav v_\ks \\ \nablav \psi_\ks \end{pmatrix}_\dyn = - \Lv_\ks^{-1} \begin{pmatrix} \jv_n \\ \jv_q \end{pmatrix} ~.
\ee
Subtracting the first equation from the second we find that the dynamical xc fields
\be
  \begin{pmatrix} \nablav \bar{v}_\xc^\dyn \\ \nablav \bar{\psi}_\xc^\dyn \end{pmatrix} = 
  \left[ \Lv^{-1} - \Lv_\ks^{-1} \right] \begin{pmatrix} \jv_n \\ \jv_q \end{pmatrix} ~.
\ee
It is well known that the static uniform electrical conductivity $\sigma$ of a uniform electron gas tends to infinity, due to translational invariance (alias momentum conservation).  Thus we could set $1/\sigma=0$.  A better approximation, which will allow us to retain viscosity corrections to the electrical resistance, is to retain nonlocal corrections to the conductivity.  As shown in \onlinecite{UllrichVignale:02}, this amounts to approximating $\frac{1}{\sigma} \simeq  -\frac{1}{n}\nablav\eta_\xc\nablav\frac{1}{n}$ where $\eta_\xc$ is an effective viscosity coefficient, which actually represents the additional viscosity created by electron-electron interactions. To avoid misunderstandings, we emphasize that the viscosity of a Fermi liquid diverges in the limit of vanishing interaction strength, signaling a breakdown of the hydrodynamic regime. It is only in the presence of frequent interactions that a local viscosity coefficient can be defined, and this is the coefficient that appears in the above approximation.  For a discussion of the difficulties involved in calculating $\eta_\xc$ from first principles, see \onlinecite{RoyDiVentra:11,PrincipiPolini:16}.  On the other hand, the thermal conductivity of an interacting electron gas is finite, but it diverges in the non-interacting electron gas~\cite{PrincipiVignale:15}.  Therefore, terms containing $1/\kappa$ can be set to zero in the non-interacting matrix $\Lv_\ks^{-1}$, but remain finite in $\Lv^{-1}$. 

Putting all the pieces together we finally obtain our effective exchange-correlation fields in the following form:
\be
  \begin{pmatrix} \nablav \bar{v}_\xc^\dyn \\[1em] \nablav \bar{\psi}_\xc^\dyn \end{pmatrix} = 
  \begin{pmatrix} -\frac{e}{n}\nablav\eta_\xc\nablav\frac{e}{n}+\frac{e^2 \Pi^2}{\kappa T} & \frac{e \Pi}{\kappa T} \\[1em]
  \frac{e \Pi}{\kappa T} & \frac{1}{\kappa T}\end{pmatrix}
  \begin{pmatrix} \jv_n \\[1em] \jv_q \end{pmatrix} ~. \label{Dynxcfields}
\ee
As mentioned already in Sec.\ \ref{SEC:TDthermalTransport} these xc potentials must be added to the ALDA xc potentials to generate the full xc potentials. Using the linear response relations \eqref{jnjq_VT} implies that we are expressing electric and thermoelectric \emph{fields}, $\Ev_\xc = \nabla \bar{v}_\xc^\dyn$ and $\Ev_{\thermal,\xc} = \nablav \bar{\psi}^\dyn$, respectively, and hence we are determining corrections beyond the adiabatic approximations, because the potentials are determined by the currents. It must be borne in mind that the condition for the validity of this local description of the dissipative fields is essentially the same as the condition for the validity of hydrodynamics.  In particular, for electronic systems, the temperature is required to be sufficiently high to make the electron mean free path shorter than the other relevant length scales of the system (e.g., the geometric size of the system and the scale of variation of the potential). In the presented discussion we have tacitly assumed that the effective KS particle and energy currents are identical to the currents of the interacting system. Strictly speaking this can only be justified for the longitudinal component of the currents since we are working in a \emph{density} functional and not a \emph{current density} functional framework. With this caveat, however, we can assumed the currents to be equal since the \emph{excess} energy densities, i.e., the change of the energy densities, are identical in the KS and the interacting system, since we are working only to first order the post ALDA corrections.

\section{Applications}

\subsection{Dynamical corrections to the resistivity}

The incompleteness of the Landauer-B\"uttiker approach and the existence of dynamical many-body corrections to the resistance of mesoscopic as well as extended electronic system has long been recognized, starting with the pioneering work of \cite{SaiDiVentra:05}.  Dynamical many-body corrections cannot be captured by any static mean-field potential.  Rather, they arise from time-dependent fluctuations of the effective potential in the out-of-equilibrium system--an effect that persists even in the zero-frequency limit and was interpreted, in \cite{SaiDiVentra:05,VignaleDiVentra:09,RoyDiVentra:11} as a manifestation of the viscosity of the electron liquid.

More recently in \cite{AndreevSpivak:11} a hydrodynamic theory of electric and thermal transport in strongly correlated electronic systems was formulated by Andreev and co-workers.  With  this theory they have found viscosity corrections to the resistance, which are actually of the form predicted in~\cite{SaiDiVentra:05,VignaleDiVentra:09,RoyDiVentra:11}, but also an interesting thermal correction to the resistance, which arises from the local heating of the electron gas and the additional potential difference generated by the temperature gradients via the thermopower effect.  In this section we show that thermal corrections to the dc resistivity arise naturally from post-ALDA corrections to the xc effective potentials in the framework of the thermal DFT. For one dimensional systems, these corrections have exactly the same form that was predicted in Ref.~\cite{AndreevSpivak:11}.

We begin by recalling that if $I$ is the current flowing through a conductor, then the energy dissipated per unit time is $W=RI^2$, where $R$ is the resistance of the conductor.  To calculate $R$ we perform a microscopic calculation of the dissipated power. This is the work done by the external fields on the currents -- electrical and thermal.   The dissipated power can be decomposed into a ``Kohn-Sham" part and an xc correction.  The Kohn-Sham contribution is well described by the LB formalism, while the xc contribution is responsible for the ``dynamical corrections".  The xc contribution to the dissipation reads
\be
W_\xc=\ind{3}{r}~ \overline{ \jv_n(\rv) \cdot \Ev_\xc (\rv)+\jv_q(\rv)\cdot \Ev_{\thermal, \xc}(\rv)}
\ee
where the overline denotes the time average over a period of oscillation of the fields, which tends to infinity at the end of the calculation (i.e., $\omega \to 0$).  Only the part of the effective fields that oscillates in phase with the currents contributes to the dissipation.  Therefore,  the dissipation arises entirely from the dynamical contributions to the xc fields, i.e., from the post-ALDA terms.  This is because the ALDA fields are in phase with the densities, and therefore 90$^0$ out of phase with the current densities, resulting in zero dissipation.  Taking into account the form~\eqref{Dynxcfields} of the dynamical xc fields we easily find
\be\label{Dissipation}
W_\xc = \ind{3}{r}~ \overline{ e^2 \eta_\xc \left\vert \nablav \left(\frac{\jv_n}{n}\right)\right\vert^2+\frac{1}{\kappa T} \left\vert \jv_q(\rv)+e \Pi \jv_n(\rv)\right\vert^2}~,
\ee
where both $\eta_\xc$ and $\kappa$ have a spatial variation (not explicitly shown) due to their dependence on the local density.

Notice that the dissipation vanishes in the case of a uniform electron gas moving with constant velocity $\vv$, for which  $\jv_n/n = \vv$ is uniform (vanishing gradient) and $\jv_q=- e \Pi \jv_n$. As shown in Eq.\ \eqref{UniformCurrents} in Sec.\ \ref{SEC:TMpotential} this simply corresponds to a Galilean transformation to reference frame moving with a constant velocity $\vv$, which should not generate dissipation. On the other hand, uniform currents injected in a non-uniform electron gas do generate dissipation.  

In order to proceed, let us consider a quasi-one dimensional conductor of length $L$ and cross sectional area $a^2$, so that its volume is $V=La^2$.  The current $I$ is related to the current density $j_n$ by $I=-ej_na^2$.  Hence the dissipation can be written as 
\be\label{ResistivityDefined}
RI^2 = e^2 R |\jv_n|^2 a^4 =   e^2 \rho |\jv_n|^2  V\,,
\ee 
where we have introduced an effective resistivity $\rho$ such that $R=\rho L/a^2$. Note that this definition is purely formal and does not imply that the metal is a ohmic conductor.

The first term in the integrand of Eq.~(\ref{Dissipation}) is now rewritten as
\be
W_\xc^\visc =  \left\langle \eta_\xc \left \vert \frac{\nablav n}{n^2}\right\vert^2 \right\rangle e^2|\jv_n|^2 V 
\ee
where the angular brackets denote a spatial average and we have made use of the constancy of $|\jv_n|$ to pull it out of the average.  Comparison with Eq.~(\ref{ResistivityDefined}) then leads to the well-known form of the viscous contribution to the dynamical resistivity~\cite{SaiDiVentra:05,VignaleDiVentra:09}:
\be
\rho_\xc^\visc =  \left\langle \eta_\xc \left \vert \frac{\nablav n}{n^2}\right\vert^2 \right\rangle\,.
\ee
Assuming the system to be weakly inhomogeneous, i.e., working to second order in $|\nablav n|$, we replace $\eta_\xc$ by its spatial average $\langle \eta_\xc\rangle=\bar\eta_\xc \hbar \langle n \rangle / e^2$, where $\bar \eta_\xc$ is dimensionless and $\langle n \rangle$  is the average density. Thus, we arrive at the compact formula
\be\label{CompactFormula}
\rho_\xc^\visc \simeq \frac{\hbar a_\eff}{e^2}\frac{\bar \eta_\xc}{\langle n\rangle a_\eff^3} \,,~~~~\frac{1}{a_\eff^2} \equiv  \frac{\langle|\nablav n|^2 \rangle}{\langle n \rangle ^2}\,.
\ee
Notice that $a_\eff$ plays the role of  a characteristic length scale for density variations.  The validity of the hydrodynamic description  requires that the mean free path between electron-electron collisions, given by the Fermi velocity times the quasiparticle lifetime (see~\onlinecite{GiulianiVignale:05}, Eq. 8.93), $\lambda \simeq (E_\fermi / \kB T)^2 k_\fermi^{-1}$, where $k_\fermi$ is the Fermi wave vector and $E_\fermi$ is the Fermi energy, remains smaller than $a_\eff$. Thus, the temperature cannot be too low.  In particular, the order of magnitude of the dimensionless viscosity $\bar \eta_\xc$ is $[E_\fermi/(\kB T)]^2$ (see \cite{RoyDiVentra:11}), which is typically somewhat larger, but not much larger than $1$.  In the homogeneous limit $a_\eff \to \infty$ and the dynamical correction vanishes, at any given temperature.

We now show that the second term of Eq.~(\ref{Dissipation}) yields the thermal corrections first identified in ~Ref.\ \cite{AndreevSpivak:11}. Due to the local conservation laws [cf.\ Eq.\ \eqref{heatConservation} in Sec.\ \ref{SEC:TMpotential}] both $\jv_n$ and $\jv_q$ are spatially constant in a one dimensional system. The steady heat current $\jv_q$ is determined by the condition that the spatial average of the gradient of the temperature vanishes. Accordingly $\jv_q$ is given in terms of $\jv_n$ by the requirement
\be
e \left\langle \frac{\Pi}{\kappa}\right\rangle \jv_n+\left\langle \frac{1}{\kappa}\right\rangle \jv_q=0 ~.
\ee
from which we deduce
\be
\jv_q= - e\frac{\langle\Pi\kappa^{-1}\rangle}{\langle\kappa^{-1}\rangle} \jv_n ~.
\ee
 Inserting this in Eq.~(\ref{Dissipation}) for the dissipation we get
 \be
 W_\xc^\thermal= \left\langle \frac{1}{\kappa T}\left\vert \Pi - \frac{\langle\Pi\kappa^{-1}\rangle}{\langle\kappa^{-1}\rangle} \right\vert^2 \right\rangle  e^2 |\jv_n|^2 V ~.
\ee
As mentioned earlier the Peltier coefficient is $\Pi=- \kB T s / e$, where $s$ is the entropy per particle (see Eq.~\eqref{UniformCurrents}). Hence we obtain
\be
 W_\xc^\thermal= \kB^2 T \frac{\left\langle \delta s^2 \right\rangle}{\left\langle \kappa \right\rangle} |\jv_n|^2 V ~,
\ee
where we defined
\begin{align}
  \left\langle\delta s^2\right\rangle & = 
  \left\langle \kappa^{-1} \right\rangle \left\langle s^2 \kappa^{-1} \right\rangle 
  - \left\langle s \kappa^{-1} \right\rangle^2~.
\end{align}
This result agrees with the thermal contribution to the resistivity reported in Eq. (4) of Ref.~\onlinecite{AndreevSpivak:11}. Now, assuming again weak inhomogeneity, we can approximate $s(n) \approx s(\langle n \rangle) + 
s'(\langle n \rangle) a_\eff \nablav n$, where $a_\eff$ is the characteristic length scale of the inhomogeneity, defined in Eq.~\eqref{CompactFormula}. This leads to
\begin{align}
  \left\langle\delta s^2\right\rangle & \approx \left[ s'(\langle n \rangle) \right]^2
  \left\langle |\nablav n|^2 \right\rangle a_\eff^2
\end{align}
Finally we approximate the entropy with the electron gas formula $s \propto \kB T/E_\fermi$, where $E_F \propto n^{2/3}$, which implies $s'=-2 s/3n$. Accordingly we have
\begin{align}
  \left\langle\delta s^2\right\rangle & \propto \left(\frac{\kB T}{E_\fermi}\right)^2
\end{align} 
Comparing with Eq.~(\ref{ResistivityDefined})  yields the thermal correction to the resistivity
\be
  \rho_\xc^\thermal = \frac{\kB^2 T}{e^2}
  \frac{\left\langle \delta s^2 \right\rangle}{\left\langle \kappa \right\rangle} 
  \propto \frac{\hbar a_\eff}{e^2}
  \frac{\kB^2 T}{\left\langle \kappa \right\rangle\hbar a_\eff}
  \left(\frac{\kB T}{E_\fermi}\right)^2 ~.
\ee
Thus the ratio of the thermal correction to the viscous corrections is
\be
\frac{\rho_\xc^\thermal}{\rho_\xc^\visc} \propto \frac{k_B^2T \langle n \rangle a_\eff^2}{\hbar \langle \kappa \rangle \bar \eta_\xc}
\left(\frac{\kB T}{E_\fermi}\right)^2
\propto \frac{\hbar \kB \langle n \rangle}{m \langle \kappa \rangle} \frac{\left[ \langle n \rangle^{1/3} a_\eff \right]^2}{\bar \eta_\xc} 
\left(\frac{\kB T}{E_\fermi}\right)^3~.
\ee
Taking $\langle n \rangle^{1/3} a_\eff \simeq \bar \eta_\xc \simeq \frac{k_BT}{E_F} \simeq 1$ we find that both resistivities are of the same order for a thermal conductivity of the order of $\kappa \simeq 10^2 \mathrm{W} (\mathrm{mK})^{-1}$ and a density of the order of $n \simeq 10^{29} \mathrm{m}^{-3}$.

\subsection{Temperature oscillations and transient heat currents in nanoscale conductors}
Thermal DFT allows for a efficient description of thermoelectric transport in nanoscale junctions, because the charge and heat transport of the interacting system is mapped onto a non-interacting effective system. This means that not only the current--voltage characteristic, but the thermoelectric transport properties can--in principle--be obtained from the Landauer-B\"uttiker formalism generalized to leads at different chemical potentials and thermomechanical potentials~\cite{EichVignale:14b}. A typical transport setup is sketched in Fig.\ \ref{FIG:transportSetup}, Sec.\ \ref{SEC:TMpotential}. The \emph{steady-state} particle current, $I$, and heat current, $Q$, in lead $\alpha$ are given by
\begin{subequations} \label{steadyStateCurrents}
  \begin{align}
    I_{\alpha} & = \frac{1}{\hbar} \sum_{\alpha'} \frac{1}{2 \pi} \indll{\epsilon}{-\infty}{\infty} T_{\alpha \alpha'}(\epsilon) \left(f_{\alpha} - f_{\alpha'}\right) ~, \label{ParticleCurrent} \\
    Q_{\alpha} & = \frac{1}{\hbar} \sum_{\alpha'} \frac{1}{2 \pi} \indll{\epsilon}{-\infty}{\infty} \epsilon T_{\alpha \alpha'}(\epsilon) \left(f_{\alpha} - f_{\alpha'}\right) ~, \label{HeatCurrent}
  \end{align}
\end{subequations}
where the $f_\alpha = f(\beta_\alpha(\epsilon - \mu_\alpha))$ are the Fermi-Dirac occupation functions determined by the (inverse) temperature, $\beta_\alpha = (\kB T_\alpha)^{-1}$, and chemical potential, $\mu_\alpha$, in the corresponding lead and $T_{\alpha \alpha'}(\epsilon)$ is the energy dependent transmission between leads $\alpha$ and $\alpha'$. 

\begin{figure}
  \begin{center}
    \includegraphics[width=.5\textwidth]{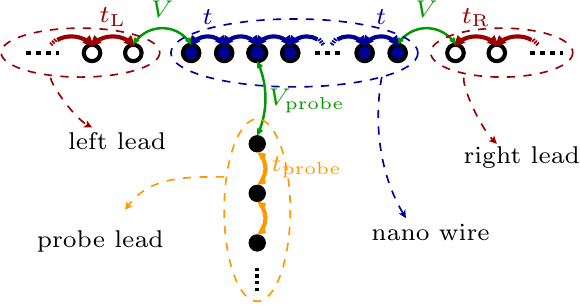}
  \end{center}
  \caption{Tight binding model of a transport setup for the determination of a local temperature and chemical potential. A uniform wire is model by a finite tight binding chain. The left and right leads are semi-infinite tight-binding chains which are coupled with the wire and are allowed to exchange particles and energy with the nano wire. Another semi-infinite tight binding chain, which is taken to be weakly coupled to the wire is treated as a local probe by imposing that no particles and energy are exchanged with the wire. Sweeping this probe lead over the wire while imposing the zero-currents conditions maps out the local temperature and chemical potential.}
  \label{FIG:localTemperatureSetup}
\end{figure}
As already mentioned in Sec.\ \ref{SEC:TMpotential}, leads can have two distinct roles: 1) They can be sources or sinks for charge and energy, which means they are used to drive the system out of equilibrium. 2) They can measure the system, e.g., by adjusting their temperature and chemical potential in order to suppress any heat or charge current into the lead in the steady-state limit. Leads which are subject to the ``zero-currents'' condition are sometimes referred to as floating probes~\cite{EngquistAnderson:81,DiVentra:08}. Since there is no particle and energy flow they can be viewed as being in local equilibrium with the part of the system to which they are connected. However, these leads affect the transmission between other leads. One way to minimize this unwanted perturbation is to couple them very weakly to the system. It has been shown that under the weak coupling assumption the ``zero-currents'' condition for the probe lead can be written as the requirement that the steady state value of the particle and energy densities of the part of the system that is coupled to the probing lead can be written as an equilibrium expectation value~\cite{EichVignale:14b,Stafford:16}. Denoting these local densities as $n_i$ and $h_i$, respectively, the conditions which determine the local temperature and potential explicitly read
\begin{subequations} \label{localPotentials}
  \begin{align}
    n_{i} & = \frac{1}{2 \pi} \indll{\epsilon}{-\infty}{\infty} D_{i}(\epsilon) f_{i} ~, \label{localDensity} \\
    h_{i} & = \frac{1}{2 \pi} \indll{\epsilon}{-\infty}{\infty} \epsilon D_{i}(\epsilon) f_{i} ~, \label{localEnergy}
  \end{align}
\end{subequations}
where the effective local density of state, $D_i(\epsilon)$, and the densities $n_i$ and $h_i$ are determined by the structure of the device and the other leads. The (inverse) temperature, $\beta_i$, and chemical potential $\mu_i$, which, in turn, yield the local occupation function, $f_i = f(\beta_i(\epsilon - \mu_i))$, have to be adjusted such that conditions \eqref{localPotentials} hold.

\begin{figure}
  \begin{center}
    \includegraphics[width=.5\textwidth]{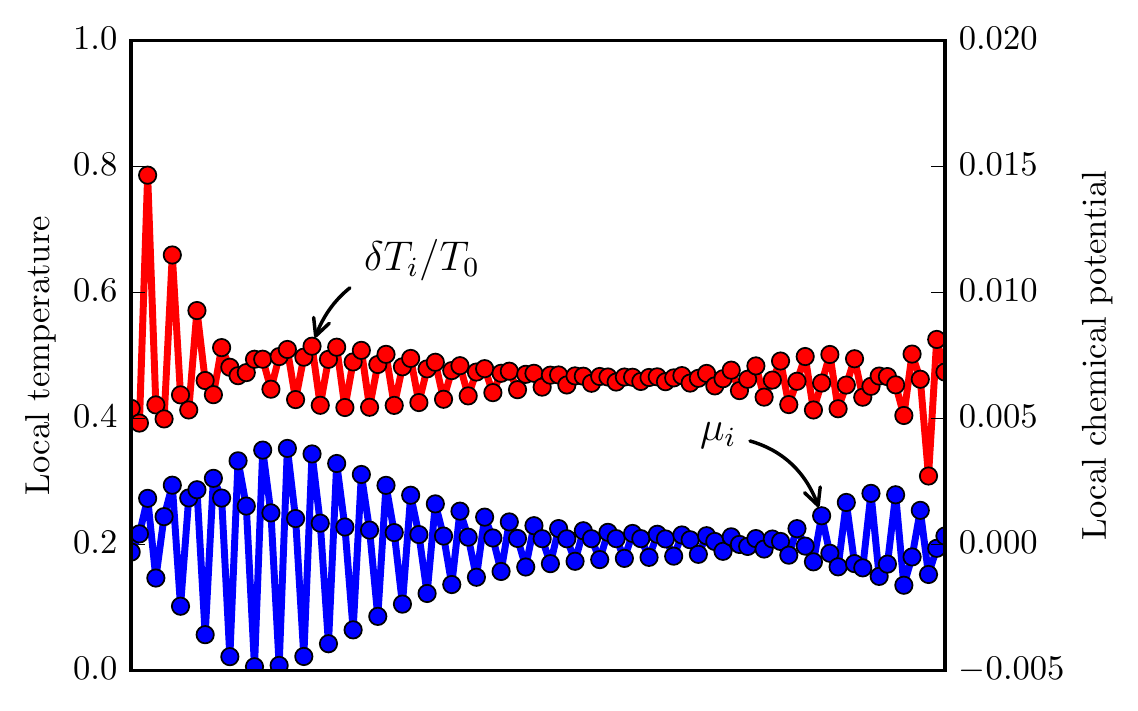}
  \end{center}
  \caption{Plot showing the oscillations of the local temperature and the chemical potential in a metallic nano wire exposed to a temperature gradient. Both, $T_i$ and $\mu_i$ oscillate with a periodicity of $\lambda = 1 / 2 k_\fermi$. The energy scale is chosen to be the hopping $V$ from the wire into the leads. The hopping in the wire is $t=V$ and in the leads we have set $t_\mathrm{R}=t_\mathrm{L}=2V$. The reference temperature corresponds to $\kB T_0 = 0.025V$. The temperature in the left lead is doubled by applying a thermomechanical potential $\psi=1$. The meaning of the parameters can be inferred from Fig.\ \ref{FIG:localTemperatureSetup}}
  \label{FIG:localOscillations}
\end{figure}

The definition of a local temperature has recently attracted a lot of attention due to the fact that nowadays experiments achieve a spatial temperature resolution down to the nanometer scale, e.g., via scanning thermal microscopy~\cite{Majumdar:99,YuKim:11,KimLee:11,KimReddy:12,MengesGotsmann:12} or transmission electron microscopy~\cite{MecklenburgRegan:15}. A detailed discussion of the various theoretical approaches~\cite{CasoLozano:10,DubiDiVentra:11,BergfieldStafford:13,EichVignale:14b,Stafford:16} is beyond the scope of this work. Here, we highlight how the zero-current conditions \eqref{localPotentials} can be used to study the local temperature and chemical potential oscillations in a conducting nano wire~\cite{EichVignale:16}. A pictorial representation of the tight-binding Hamiltonian describing the nano wire suspended between two leads, together with a  third ``probe'' lead to determine the local chemical potential and temperature, is shown in Fig.\ \ref{FIG:localTemperatureSetup}. If we heat up the nano wire on one end by applying a thermomechanical potential a steady-state current will form after a characteristic time determined by the inverse of the decay rate of the electrons from the wire into the leads. In this long time limit also the energy density becomes stationary and yields, by virtue of Eqs.\ \eqref{localPotentials}, the local temperature in the wire as measured by the weakly coupled probe lead. In Fig.\ \ref{FIG:localOscillations} we show a typical temperature profile for a wire modeled by $100$ tight-binding sites. At low temperatures the local temperature exhibits characteristic $2 k_\fermi$ Friedel oscillations as pointed out in Refs.\ \onlinecite{DubiDiVentra:09,BergfieldDiVentra:15,EichVignale:16}.

Finally we address the transient currents. One advantage of the approach to thermoelectric transport using the thermomechanical potential is that we can work in the so-called unpartitioned approach~\cite{Cini:80,StefanucciAlmbladh:04,EichVignale:14b}. The usual way of addressing different temperatures in the LB or Meir-Wingreen formalism~\cite{MeirWingreen:92} is to decompose the system into various parts. This means that there is no coupling between these subsystems initially, which allows to equilibrate the subsystems at different temperatures. The dynamics is then triggered by establishing the coupling between the subsystems. The thermomechanical potential, however, mimics spatially varying temperatures directly and therefore does not rely on the decomposition of the system into uncoupled parts. As already mentioned in Sec.\ \ref{SEC:TMpotential} this does not have any effect on the steady state--at least in non-interacting systems. The transient currents, however, will in general be different.

\begin{figure}
  \begin{center}
    \includegraphics[width=.5\textwidth]{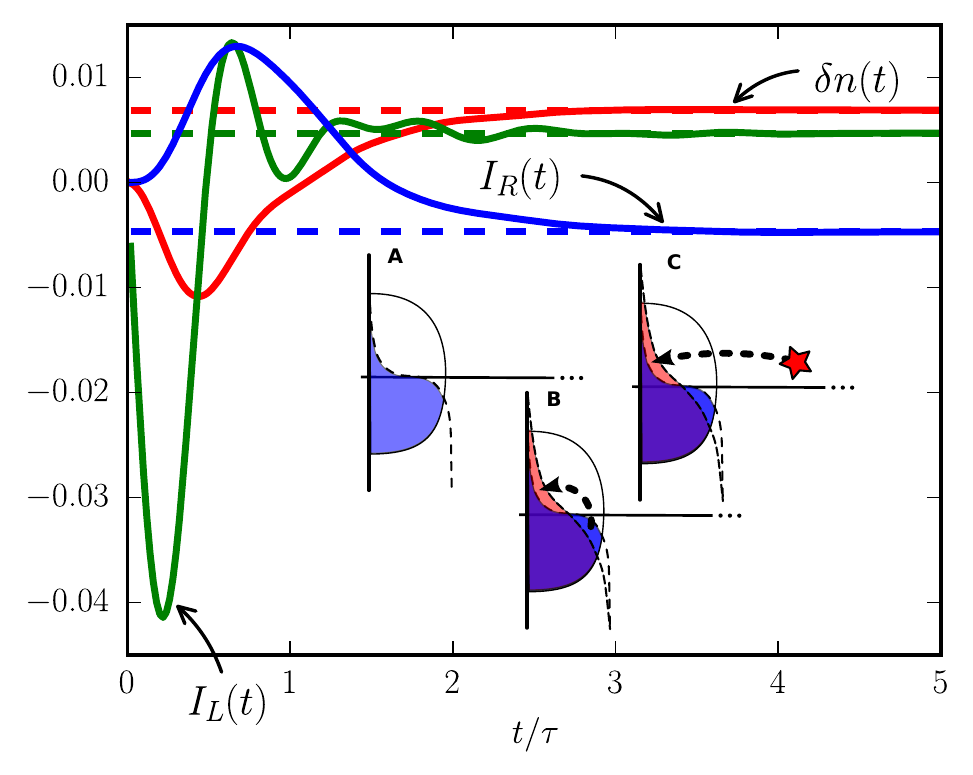}
  \end{center}
  \caption{Transient particle current trough a quantum dot modeled by a single impurity site coupled to two metallic leads. The energy scale is set by the coupling to the leads $V$. As in the system shown in Fig.\ \ref{FIG:localOscillations} the lead hopping is $t_\mathrm{R}=t_\mathrm{L}=2V$. The dispersion of both leads is shifted by $-1V$, which leads to a quasi-particle energy of the dot slightly above the chemical potential. The coupled charge and heat dynamics are triggered by doubling the temperature in the left lead. In the inset we sketch the initial population of the left lead (A), and the population after raising the temperature (B). The repopulation of the electrons from below to above the chemical potential is initially facilitated by the quantum dot (C), which leads to an initial flow of electrons from the dot into the left lead.}
  \label{FIG:TransientCharge}
\end{figure}

In Fig.\ \ref{FIG:TransientCharge} we show the transient charge current through a quantum dot. Interestingly we find that the initial current is opposite to the steady state current. Since we are inducing the charge current by a temperature gradient this can be viewed as a sign change of the Seebeck coefficient at short times (high frequencies). This initial reversal of the Seebeck effect can be understood as a quantum dot assisted population transfer in the left lead (cf.\ inset in Fig.\ \ref{FIG:TransientCharge}): Electrons have to move from below to above the chemical potential due to the increase in temperature. The impurity, which is positioned above the chemical potential, can support this transfer of occupation by providing electrons. Afterwards the impurity is resupplied by electrons from the right lead, since also the impurity needs to adapt to the higher temperature to its left. Then in the steady-state the heat flows in the ``natural'' direction from hot to cold. This means that electrons below the chemical potential flow from the colder to the hotter lead, and electrons above the chemical potential in the opposite direction. Since the impurity site is located above the chemical potential the latter direction is preferred and we get a net particle current in the direction of the heat flow.

\begin{figure}
  \begin{center}
    \includegraphics[width=.7\textwidth]{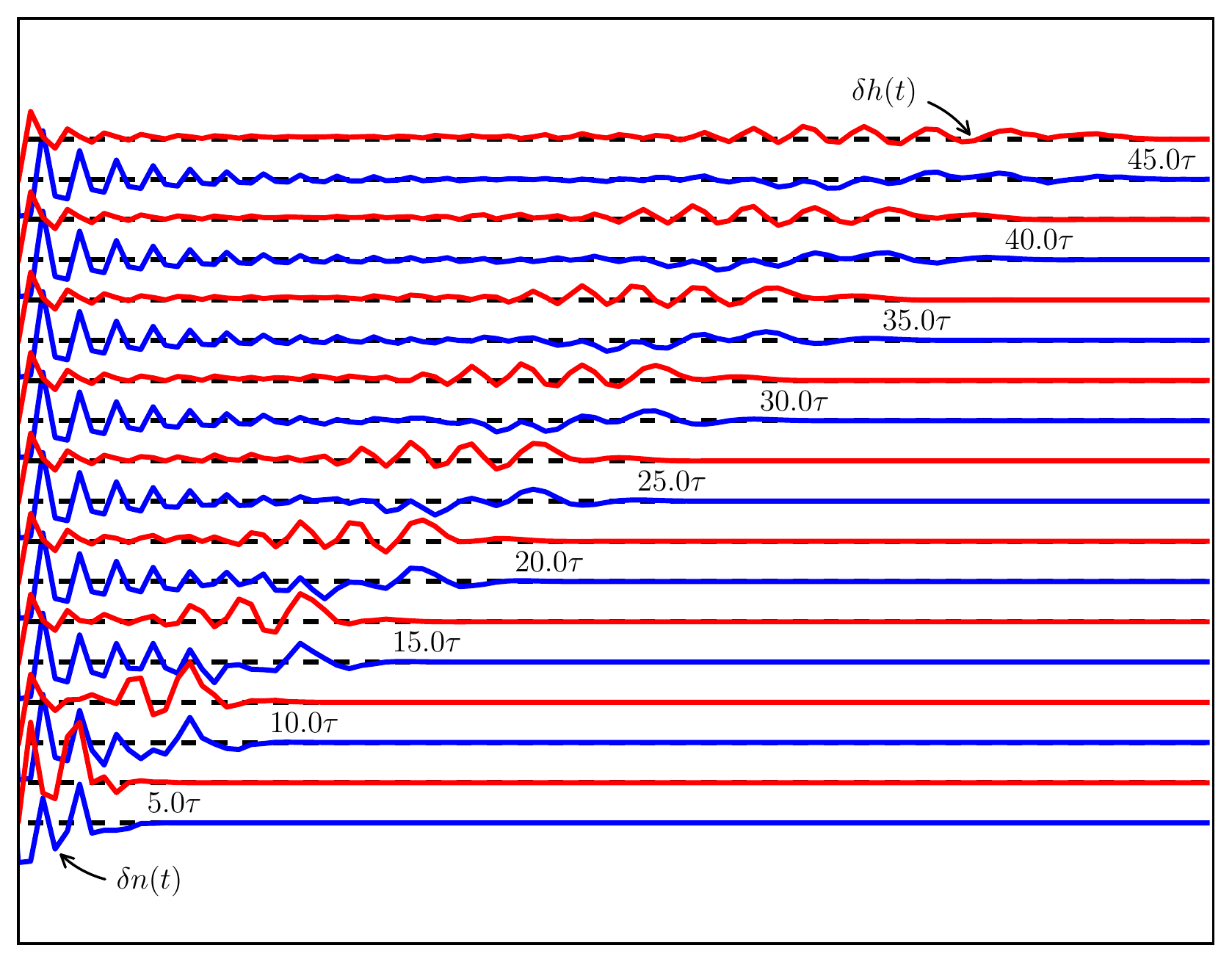}
  \end{center}
  \caption{Initial wave fronts of the density (blue), $\delta n$, and energy density (red), $\delta h$, variation. For a nano wire modeled by $100$ tight-binding sites (cf.\ Fig.\ \ref{FIG:localOscillations} for steady state). The velocity of the charge and heat wave fronts are identical for this non-interacting model, which is due to the fact that, both, charge and energy are carried by the electrons leading to a Wiedemann-Franz-like behavior. The plot shows snapshots of the density- and energy-variation profiles along the wire ($x$ axis) at intervals of $5 \tau$ before the wave fronts get reflected at the surface to the right lead.}
  \label{FIG:TransientWaves}
\end{figure}

Lastly, we turn to the the transients in the nano wire. In Fig.\ \ref{FIG:TransientWaves} we show the wave fronts of the charge and energy waves in a nano wire. While in Fig.\ \ref{FIG:localOscillations} we showed the steady state local temperature and chemical potential profile, in Fig.\ \ref{FIG:TransientWaves} we focus on the initial propagation of the charge and energy waves. Fig.\ \ref{FIG:TransientWaves}  depicts snapshots of the density and energy profile in the nano wire before the wave fronts hit the boundary to the left lead and are partially reflected. In this simple non-interacting model, charge and energy perturbations travel at the same speed through the nano wire, because both are carried by electrons. In Ref.\ \onlinecite{EichVignale:16} we have found that the velocity of the initial wave fronts only depends on the hopping amplitude, $t$, in the nano wire and not on the density or global reference temperature.

\section{Summary and Outlook} \label{SEC:Summary}

In this work we took stock of recent developments to address thermoelectric transport within a density functional framework. Starting from the fundamental equations of hydrodynamics a novel DFT, dubbed {\it thermal DFT}, has been recently proposed~\cite{EichVignale:14a,EichVignale:14b}. The gist of thermal DFT is to include the energy density as fundamental variable, thereby allowing for a direct description of combined heat and charge transport. The foundation for the corresponding density functionalization is based on the thermomechanical potential, a concept introduced by Luttinger already half a century ago~\cite{Luttinger:64a}. We have presented a thorough discussion and physical motivation for this potential. Furthermore, we have provided, for the first time, the microscopic equation of motion for the energy density in the presence of the aforementioned thermomechanical potential, which allows to directly connect the microscopic theory to the hydrodynamic equations. We have then proceeded to discuss the formal construction of thermal DFT in great detail, filling gaps left in earlier publications. For example, we presented the static limit of thermal DFT highlighting its relation to standard finite temperature DFT, and presented an invertibility argument for the fully time-dependent theory. We have compared the thermal DFT approach to other density functional approaches to thermoelectric transport based on the stochastic Sch\"odinger equation. In passing we have pointed out that thermal DFT, at present, is the only approach which directly addresses the energy or heat degrees of freedom. Moreover, we have explicitly derived the adiabatic local-density approximation, paving the way for an implementation of the thermal DFT framework. Furthermore, we have addressed the construction of so-called dynamical corrections, and emphasized the connection to other works promoting a hydrodynamic picture for interacting electrons. Finally, we have compiled some recent results on thermal transport employing the Luttinger's thermomechanical potential, e.g., the calculation of local temperatures which can be directly related to recent experiments.

An open challenge remains the inclusion of electron-phonon, or generally fermion-boson interactions, in the framework of thermal DFT. In principle there are two contributions to the thermal resistivity for mesoscopic electronic devices: a contribution due to the electronic degrees of freedom and a contribution due to other degrees of freedom, e.g., ionic motion in the form of molecular vibrations or phonons. In this review we have only focused on the electronic contribution, which means that we are addressing inhomogeneous systems in the hydrodynamic regime, where electron-electron collisions dominate over scattering with impurities and/or phonons. However, attempts have been already made to include electron--impurity scattering in a time-dependent DFT framework~\cite{UllrichVignale:02}, which can be generalized in order to address thermoelectric transport. Furthermore the combination of a stochastic density functional approach, where the bosonic degrees of freedom are treated as a randomizing bath, with thermal DFT is a promising route to include fermion-boson coupling. 

Another scenario in which fermion-boson coupling may be neglected completely is the field of atomtronics~\cite{SeamanHolland:07}, where electronic devices are simulated by cold atoms. In recent years transport experiments with cold atomic gases have received a lot of attention~\cite{BrudererBelzig:12,BrantutEsslinger:12,BrantutGeorges:13}, since they can be used to simulate purely fermionic transport devices~\cite{ChienDiVentra:15} and may prove to be the most relevant experiments which can be directly addressed with thermal DFT.

We believe that thermal DFT provides an important step towards generalized hydrodynamics at the microscopic scale. While there are many open issues and challenges, we feel the most urgent one is to include the corrections due to electron-electron interactions into thermoelectric transport calculations.  This will not only lead to a better understanding of the importance of such corrections, but may lead to a direct numerical approach for solving the microscopic hydrodynamic equations, bypassing the need for solving the effective Kohn-Sham equations. This opens the door for large scale computations for thermoelectric devices.

\begin{acknowledgments}
  We gratefully acknowledge support from the Deutsche Forschungsgemeinschaft under DFG Grant No. EI 1014/1-1 (F. G. E.), and the DOE under Grants No. DE-FG02-05ER46203 (G. V.) and DE-FG02-05ER46204 (M. D.).
\end{acknowledgments}

\begin{appendix}
  \section{Continuity equation for the energy density} \label{APP:eom}
  In this appendix we supply details of the derivation of the continuity equation for the energy density. Even though the equations of motion for the energy density have been reported in the literature~\cite{Langer:62}, usually the explicit expressions are only given for the kinetic contribution~\cite{QinShi:11}, or in the absence of the thermomechanical potential~\cite{PuffGillis:68}, or considering only weakly inhomogeneous systems~\cite{Luttinger:64a}. Here we provide for the first time--to the best of our knowledge--the continuity equation for the energy density including a nontrivial thermomechanical potential. The Heisenberg equation of motion is
  \be\label{Heisenberg0}
    \partial_t \hat h(\rv) = \frac{i}{\hbar}[\hat H,\hat h(\rv)]
    = \frac{i}{\hbar} \ind{3}{r'} \{1+\psi(\rv')\}[\hat h_0(\rv'), \hat h_0(\rv)]\{1+\psi(\rv)\}
  \ee
  or, in a more symmetric form,
  \be\label{Heisenberg}
    \partial_t \hat h(\rv) = \frac{i}{\hbar}\ind{3}{r'} \ind{3}{r''} \{1+\psi(\rv')\}[\hat h_0(\rv'), \hat h_0(\rv'')]\{1+\psi(\rv'')\}\delta(\rv''-\rv)\,,
  \ee
  where $\hat h_0(\rv)$ is the energy density operator \emph{in the absence} of Luttinger's thermomechanical potential. The core of the task is to calculate the commutator $[\hat h_0(\rv'), \hat h_0(\rv'')]$.  The commutator can be split into three parts corresponding to the kinetic energy, $\hat t_0$,  the external potential energy, $\hat v_0$, and the interaction energy, $\hat u_0$:
  \be
    [\hat h_0(\rv'), \hat h_0(\rv'')] =  [\hat t_0(\rv'), \hat t_0(\rv'')] +\left\{[\hat t_0(\rv'), \hat v_0(\rv'')]+[\hat v_0(\rv'), \hat t_0(\rv'')]\right\}+\left\{[\hat t_0(\rv'), \hat u_0(\rv'')]+[\hat u_0(\rv'), \hat t_0(\rv'')]\right\}\,,
  \ee
  where we have discarded terms such as $[\hat v_0(\rv'),\hat v_0(\rv'')]$ which are zero, since they only include density operators. 

  \subsection{Kinetic energy} \label{APP:eomKin}
  Adopting the positive definite form, $t_2$ [cf.\ Eq.\ \eqref{t2} in Sec.\ \ref{SEC:TMpotential}], for the kinetic energy we have
  \be
    \frac{i}{\hbar}[\hat t_0(\rv'),\hat t_0(\rv'')]=\frac{i\hbar^3}{(2m)^2} \left[\left(\partial_{i'}{\hat\Phi^\dagger}(\rv')\right) \left(\partial_{i'}{\hat\Phi (\rv')}\right),\left(\partial_{j''}{\hat\Phi^\dagger (\rv'')}\right) \left(\partial_{j''}{\hat\Phi}(\rv'')\right)\right]
  \ee
  where $\partial_{i'}$ is the derivative with respect to $r_i'$, where $\partial_{j''}$ is the derivative with respect to $r_j''$, and a sum over repeated indices is implied.  The commutator is readily evaluated and the result is
  \be
    \frac{i}{\hbar}[\hat t_0(\rv'),\hat t_0(\rv'')]
    = \frac{i\hbar^3}{(2m)^2} \left\{\left(\partial_{i'}{\hat\Phi^\dagger}(\rv')\right)\left(\partial_{j''}{\hat\Phi}(\rv'')\right)
    - \left(\partial_{j''}{\hat\Phi^\dagger (\rv'')}\right)
    \left(\partial_{i'}{\hat\Phi (\rv')}\right) \right\}\left(\partial_{i'}\partial_{j''}\delta(\rv'-\rv'')\right)
  \ee
  In order to accurately handle the second derivative of the $\delta$-function distribution we multiply the above expression by ``test functions" $\phi(\rv')$ and $\chi(\rv'')$ and integrate over $\rv'$ and $\rv''$. Because of the antisymmetry of the commutator, only the antisymmetric component of the product $\phi\chi$, i.e.,   $[\phi(\rv')\chi(\rv'')-\phi(\rv'')\chi(\rv')] /2$ contributes to the integral.  Integrating by parts we easily find
  \ber\label{CTT}
    \frac{i}{\hbar}\ind{3}{r'}\ind{3}{r''} [\hat t_0(\rv'),\hat t_0(\rv'')]\phi(\rv')\chi(\rv'')
    = \frac{i\hbar^3}{(2m)^2}\ind{3}{x} \left\{\left(\partial_{i} (\partial_i{\hat\Phi^\dagger})\phi\right)
    \left(\partial_{j} (\partial_j{\hat\Phi})\chi\right)  - \phi \leftrightarrow \chi\right\}
  \eer
  where  all the quantities on the right hand side are evaluated at the same position $\xv$.  In evaluating these expressions it must be kept in mind that derivatives enclosed within brackets only act on the functions {\it within} the brackets.   It is immediately evident that the terms in which the functions $\phi$ and $\chi$ are not differentiated will cancel out, due to antisymmetry under the interchange $\phi \leftrightarrow\chi$. 

  Allowing the derivative of only {\it one} of the functions  $\phi$ or  $\chi$ gives
  \be\label{Onederivative}
    (i)~~~~\frac{i\hbar^3}{(2m)^2}\ind{3}{x} \left\{ (\partial_i \hat\Phi^\dagger)(\partial_j\partial_j\hat\Phi)-(\partial_j\partial_j \hat\Phi^\dagger) (\partial_i\hat\Phi)\right\}\left\{ (\partial_i\phi) \chi - \phi(\partial_i\chi)\right\}
  \ee
  to which we must add the term in which {\it both} $\phi$ and $\chi$ are differentiated., i.e., 
  \be\label{Twoderivatives}
    (ii)~~~~\frac{i\hbar^3}{2(2m)^2}\ind{3}{x} \left\{ (\partial_i{\hat\Phi^\dagger})(\partial_j{\hat\Phi})-(\partial_j{\hat\Phi^\dagger}) (\partial_i{\hat\Phi})\right\}\left\{ (\partial_i\phi)(\partial_j\chi)-(\partial_j\phi) (\partial_i\chi)\right\}
  \ee
  It is convenient, at this point, to integrate by parts, with respect to $\partial_j$, {\it one half} of term (i).  One of the integrated terms exactly cancels (ii), another vanishes by antisymmetry, and the surviving one gives rise to the following elegant result
  \ber\label{CTT2}
    \frac{i}{\hbar} \ind{3}{r'}\ind{3}{r} [\hat t_2(\rv'),\hat t_2(\rv'')]\phi(\rv')\chi(\rv'')&=&-\ind{3}{x}  ~\hat\jv_{t} \cdot [(\nablav \phi) \chi-\phi (\nablav\chi)]
  \eer
  where the kinetic energy current density operator, in the absence of the Luttinger potential,  is defined as
  \be
  \left[\hat \jv_{t,0} \right]_i \equiv \frac{i \hbar^3}{8 m^2} \left\{\left[(\partial_i \partial_j \hat\Phi^\dagger)(\partial_j \hat \Phi)- (\partial_j \hat \Phi^\dagger)(\partial_i \partial_j \hat\Phi)\right]-\left[(\partial_i \hat\Phi^\dagger)(\partial_j\partial_j \hat \Phi)- (\partial_j \partial_j\hat \Phi^\dagger)(\partial_i  \hat\Phi)\right] \right\}
  \ee
  Notice that this expression can be viewed as the average of the current densities that would be naturally associated with the alternative forms of the kinetic energy density called $\hat t_1$ and $\hat t_2$ in Eqs.\ \eqref{t1} and \eqref{t2} in Sec.\ \ref{SEC:TMpotential}, respectively. Both are acceptable forms, since they differ by a divergence free field, proportional to the curl of the curl of the particle current density.  The big advantage of taking the average of the two possible definitions is that this choice leads to a simple and suggestive scaling of the energy current density in the presence of the Luttinger potential. To see this, we simply set $\phi(\rv')=1+\psi(\rv')$ and $\chi(\rv'')=[1+\psi(\rv'')]\delta(\rv''-\rv)$ in Eq.~(\ref{CTT2}).  Then, performing another integration by parts we arrive at the desired continuity equation
  \be\label{KineticContinuity}
    \partial_t \hat t(\rv)= - \nablav \cdot  \left\{[1+\psi(\rv)]^2~\hat \jv_{t0}(\rv)\right\}
  \ee
  showing that the current density associated with the $t_2$ form of the kinetic energy density  is
  \be
    \jv_{t}(\rv) = [1+\psi(\rv)]^2~\hat \jv_{t,0}(\rv)\,.
  \ee
  The two factors of $1+\psi$ have a simple physical interpretation, one comes from the rescaling of the energy density itself, the other from the rescaling of the  ``mass" that controls the velocity.
 
  \subsection{External potential energy}\label{APP:eomPot}
  Let us now consider the terms involving the potential energy density $\hat v_0(\rv)=v_0(\rv) \hat\Phi^\dagger(\rv)\hat\Phi(\rv)$.  Proceeding as before, we evaluate the integral
  \be \label{CTVCommutator}
    \frac{i}{\hbar}\ind{3}{r'}\ind{3}{r''} \left\{[\hat t_0(\rv'),\hat v_0(\rv'')]+[\hat v_0(\rv'),\hat t_0(\rv'')]\right\} \psi(\rv')\chi(\rv'')
  \ee 
  which is conveniently rewritten as
  \be 
    \frac{i\hbar}{2m}\ind{3}{r'}\ind{3}{r''}  \left[\left(\partial_{i'}\hat\Phi^\dagger(\rv')\right)\left(\partial_{i'}\hat\Phi(\rv')\right),\hat\Phi^\dagger(\rv'')\hat\Phi(\rv'')\right]v_0(\rv'')[\phi(\rv')\chi(\rv'')-\phi(\rv'')\chi(\rv')]\,.
  \ee 
  Evaluating the commutator yields
  \be\label{CTV} 
    \frac{i \hbar}{2m}\ind{3}{r'}\ind{3}{r''}  \left\{\left(\partial_{i'}\hat\Phi^\dagger(\rv')\right)\hat\Phi(\rv'')-\hat\Phi^\dagger(\rv'')\left(\partial_{i'}\hat\Phi(\rv')\right)\right\}v_0(\rv'')[\phi(\rv')\chi(\rv'')-\phi(\rv'')\chi(\rv')]\left(\partial_{i'} \delta(\rv''-\rv')\right)\,.
  \ee 
  We integrate the $\delta$ function by parts keeping in mind that only terms in which the derivative acts on $\phi$ or $\chi$ will survive.  Thus, Eq.~(\ref{CTV}) becomes
  \be
    -\frac{i \hbar}{2m}\ind{3}{x} \left\{\left(\partial_{i}\hat\Phi^\dagger \right)\hat\Phi -\hat\Phi^\dagger\left(\partial_{i}\hat\Phi\right)\right\}v_0[(\partial_i\phi)\chi-\phi(\partial_i\chi)]=-\ind{3}{x} ~ \hat\jv_{v,0} \cdot [(\nablav \phi) \chi-\phi (\nablav\chi)]\,,
  \ee
  where we have defined the potential energy current in the absence of $\psi$ potential, 
  \be
    \hat \jv_{v,0}(\rv)\equiv v_0(\rv)~\hat\jv_{n,0}(\rv)\,,
  \ee 
  and $\hat \jv_{n,0}(\rv)$ is the ordinary particle current density. Setting $\phi(\xv)=1+\psi(\xv)$ and $\chi(\xv)=[1+\psi(\xv)]\delta(\xv-\rv)$ we find that the expression~(\ref{CTVCommutator}) can be written as the divergence of a potential energy current density 
  \be
    \frac{i}{\hbar}[\hat V, \hat t(\rv)]+i[\hat T, \hat v(\rv)]=-\nablav \cdot \left\{[1+\psi(\rv)]^2 ~\hat \jv_{v, 0}(\rv)\right\}\,,
  \ee
  which yields Eq.~(\ref{JVCurrent}) of the main text. Once again, the effect of the Luttinger potential is simply to rescale the current density by $(1+\psi)^2$.

  \subsection{Interaction Energy}\label{APP:eomInt}
  The terms involving the interaction energy density, $\hat u_0(\rv)= \tfrac{1}{2} \hat\Phi^\dagger(\rv) \ind{3}{s} w(\rv-\sv) \hat\Phi^\dagger(\sv)\hat\Phi(\sv)\hat\Phi(\rv)$,  are
  \be 
    \frac{i}{\hbar}\ind{3}{r'}\ind{3}{r''} \left\{[\hat t_0(\rv'),\hat u_0(\rv'')]+[\hat u_0(\rv'),\hat t_0(\rv'')]\right\} \phi(\rv')\chi(\rv'')\,.
  \ee 
  They can be conveniently rewritten as
  \be 
    \frac{i \hbar}{4m} \ind{3}{r'}\ind{3}{r''}\ind{3}{s} ~w(\rv''-\sv)  \left[\left(\partial_{i'}\hat\Phi^\dagger(\rv')\right)\left(\partial_{i'}\hat\Phi(\rv')\right),\hat\Phi^\dagger(\rv'')\hat\Phi^\dagger(\sv)\hat\Phi(\sv)\hat\Phi(\rv'')\right][\phi(\rv')\chi(\rv'')-\phi(\rv'')\chi(\rv')]\,.
  \ee 
  Evaluating the commutator yields
  \ber
    \frac{i \hbar}{4m} \ind{3}{r'}\ind{3}{r''}\ind{3}{s}~ w(\rv''-\sv)&&\left\{\left(\partial_{i'}\hat \Phi^\dagger(\rv')\right) \hat\Phi^\dagger(\sv)\hat\Phi(\sv)\hat\Phi(\rv') -\hat\Phi^\dagger(\rv'')\hat\Phi^\dagger(\sv)\hat\Phi(\sv)\left(\partial_{i'}\hat \Phi(\rv')\right) \right\}\nonumber\\
    &&\times [\phi(\rv')\chi(\rv'')-\phi(\rv'')\chi(\rv')]\left(\partial_{i'}\delta(\rv''-\rv')\right)\nonumber\\
    +\frac{i \hbar}{4m} \ind{3}{r'}\ind{3}{r''}\ind{3}{s} ~w(\rv''-\sv)&&\left\{\hat\Phi^\dagger(\rv'') \left(\partial_{i'}\hat \Phi^\dagger(\rv')\right) \hat\Phi(\sv)\hat\Phi(\rv'') -\hat\Phi^\dagger(\rv'')\hat\Phi^\dagger(\sv)\left(\partial_{i'}\hat \Phi(\rv')\right)\hat\Phi(\rv'') \right\}\nonumber\\
    &&\times [\phi(\rv')\chi(\rv'')-\phi(\rv'')\chi(\rv')]\left(\partial_{i'}\delta(\sv-\rv')\right)
  \label{CUU} 
  \eer 
  Integrating the derivative of the $\delta$-function by parts reduces  Eq.~(\ref{CUU})  to the following form
  \ber\label{CUU2} 
    -\frac{i\hbar}{4m}\ind{3}{x}\ind{3}{s} ~w(\xv-\sv)&&\left\{\left(\partial_{i}\hat \Phi^\dagger(\xv)\right) \hat\Phi^\dagger(\sv)\hat\Phi(\sv)\hat\Phi(\xv) -\hat\Phi^\dagger(\xv)\hat\Phi^\dagger(\sv)\hat\Phi(\sv)\left(\partial_{i}\hat     \Phi(\xv)\right) \right\}\nonumber\\
    &&\times [(\partial_i\phi(\xv))\chi(\xv)-\phi(\xv)(\partial_i\chi(\xv))]\nonumber\\
    -\frac{i\hbar}{4m}\ind{3}{x}\ind{3}{s} ~[\partial_{x_i} w(\xv-\sv)]&&\left\{\hat\Phi^\dagger(\xv) \left(\partial_{s_i}\hat \Phi^\dagger(\sv)\right) \hat\Phi(\sv)\hat\Phi(\xv) -\hat\Phi^\dagger(\xv)\hat\Phi^\dagger(\sv)\left(\partial_{s_i}\hat \Phi(\sv)\right)\hat\Phi(\xv) \right\}\nonumber\\
    &&\times [\phi(\sv)\chi(\xv)-\phi(\xv)\chi(\sv)]
  \eer 
  To get the second part of  Eq.~(\ref{CUU2}) we first rewrite $\partial_{i'}\delta(\sv-\rv') = -\partial_{s_i}\delta(\sv-\rv')$, then  integrate by parts, and finally use $\partial_{s_i}w(\xv-\sv) = -\partial_{x_i}w(\xv-\sv)$. The terms in which $\partial_{s_i}$ acts on the $\hat \Phi$ fields cancel out. Note that we renamed the dummy integration variable $\rv''$ as $\xv$. 
  We now set $\phi(\xv)=1+\psi(\xv)$ and $\chi(\xv)=[1+\psi(\xv)]\delta(\xv-\rv)$ and perform  the integral over $\xv$ with the help of the $\delta$-function.
  The result from the first part Eq.~(\ref{CUU2}) can be rewritten  as 
  \be
    -\nablav_{\rv} [1+\psi(\rv)]^2~\hat \jv_{u,0}(\rv)
  \ee
  where
  \be\label{JUCurrent}
    \left[ \hat \jv_{u,0}(\rv) \right]_i \equiv  \frac{i\hbar}{4m}\ind{3}{s} w(\rv-\sv)\left\{\left(\partial_{i}\hat \Phi^\dagger(\rv)\right) \hat\Phi^\dagger(\sv)\hat\Phi(\sv)\hat\Phi(\rv) -\hat\Phi^\dagger(\rv)\hat\Phi^\dagger(\sv)\hat\Phi(\sv)\left(\partial_{i}\hat \Phi(\rv)\right) \right\}
  \ee
  is the convective part of the interaction energy current, as discussed in the main text.   Notice that this part of the current exhibits the ``standard" $(1+\psi)^2$ scaling.  The term arising from the second part of Eq.~(\ref{CUU2})is written as
  \be\label{juforce0}
    -\frac{1}{2}\ind{3}{s} [\partial_i w(\rv-\sv)]  [1+\psi(\rv)]\hat \rho_i (\rv,\sv)[1+\psi(\sv)]
  \ee
  where we have defined
  \be\label{juforce1}
    \hat \rho_i(\rv,\sv) \equiv \frac{i\hbar}{2m}\left\{\hat\Phi^\dagger(\rv)\left(\partial_{s_i}\hat \Phi^\dagger(\sv)\right) \hat\Phi(\sv)\hat\Phi(\rv) -\hat\Phi^\dagger(\rv)\hat\Phi^\dagger(\sv)\left(\partial_{s_i}\hat \Phi(\sv)\right)\hat\Phi(\rv) \right\} +(\sv \leftrightarrow \rv)\,,
  \ee 
  which is symmetric under the interchange of $\rv$ and $\sv$. Now, although this is not immediately obvious,  it turns out that Eq.~(\ref{juforce0}) can be expressed as the divergence of a power current in the following manner
  \be
    \frac{1}{2}\ind{3}{s} [\partial_i w(\rv-\sv)]  [1+\psi(\rv)]\hat \rho_i (\rv,\sv)[1+\psi(\sv)] = \nablav \cdot \hat \jv_f(\rv)
  \ee
  where
  \be\label{JFCurrent}
    \hat \jv_f(\rv) \equiv \frac {1}{4}\ind{3}{s} ~\sv\left\{[\partial_{s_i} w(\sv)] \indll{\lambda}{0}{1} [1+\psi(\rv+\lambda \sv)]\hat \rho_i (\rv+\lambda\sv,\rv+\lambda\sv-\sv)[1+\psi(\rv+\lambda\sv-\sv)]\right\}\,,
  \ee
  where $\lambda$  is a real number.  This can be directly verified by noting that $\sv \cdot \nablav$ acting on the expression within the curly brackets is equivalent to  $\partial_\lambda$.  Then
  \ber
    \nablav_{\rv}\cdot \hat \jv_f(\rv)&=&\frac {1}{4}\ind{3}{s}[\partial_{s_i} w(\sv)]  \indll{\lambda}{0}{1} \partial_\lambda
    [1+\psi(\rv+\lambda \sv)]\hat \rho_i (\rv+\lambda\sv,\rv+\lambda\sv-\sv)[1+\psi(\rv+\lambda\sv-\sv)]\nonumber\\
    &=&\frac {1}{4}\ind{3}{s}[\partial_{s_i} w(\sv)]  \Big\{[1+\psi(\rv+\sv)]\hat \rho_i (\rv+\sv,\rv)[1+\psi(\rv)]-[1+\psi(\rv)]\hat   \rho_i (\rv,\rv-\sv)[1+\psi(\rv-\sv)]\Big\}\nonumber\\
    &=& \frac {1}{2}\ind{3}{s}[\partial_{s_i} w(\sv)] [1+\psi(\rv+\sv)]\hat \rho_i (\rv+\sv,\rv)[1+\psi(\rv)]\nonumber\\
    &=& \frac{1}{2}\ind{3}{s} [\partial_i w(\rv-\sv)]  [1+\psi(\rv)]\hat \rho_i (\rv,\sv)[1+\psi(\sv)]\,.
  \eer
  To get the third line we have changed $\sv \to -\sv$ in the second term of the curly bracket in the second line, and   used the fact that $[\partial_i w(\rv)] =-[\partial_i w(-\rv)]$.  To go from the third to the fourth line just make the change of variables $\sv \to \sv -\rv$ and use the symmetry of $w$ and $\hat \rho_i$.
The final complete result for the interaction energy current is summarized by the equation
  \be
    \frac{i}{\hbar}[\hat U, \hat t(\rv)] + \frac{i}{\hbar} [\hat T, \hat u(\rv)]=-\nablav \cdot \left\{[1+\psi(\rv)]^2 ~\hat \jv_{u,0}(\rv)+\hat \jv_f(\rv)\right\}\,.
   \ee 
  Unfortunately, the nice $(1+\psi)^2$ scaling property is lost in the $\hat \jv_f$ current, which has a complicated nonlocal dependence on the $\psi$ potential.
 
  \section{Invertibility of the thermal response function} \label{APP:ThermalResponseInvertibility}
  
  In Sec.\ \ref{SEC:ActionFunctional} we have formally introduced the action functional for thermal DFT. The action functional $\tilde{\Lambda}[\tilde{v},\psi]$ generates the density, $n(\vr, t)$, and the energy density, $h(\vr, t)$, upon functional differentiation. The functional derivative of its (negative) Legendre transform, $A[n,h]$, yields the external potential, $\tilde{v}$, and thermomechanical potential, $\psi(\vr, t)$. In writing the $A[n,h]$ we tacitly assumed that we can invert relation Eq.\ \eqref{den_pot_td} for the potentials. Put differently, we assumed that there is a one--to--one correspondence between the pair of potentials $\big[\tilde{v}(\vr, t), \psi(\vr, t)\big]$ and the pair of densities $\big[n(\vr, t), h(\vr, t)\big]$. In this section we support this statement by showing that this is indeed the case at the level of linear response. The presented proof follows closely the invertibility proof of van Leeuwen~\cite{vanLeeuwen:01}, which has recently been adapted for (thermal) ensembles~\cite{Giesbertz:16,PribramJonesBurke:16}. Suppose that the system is initially in a (quasi-)equilibrium state, determined via a Hamiltonian $\hat{H}_0$, at inverse temperature $\beta$ and chemical potential $\mu$. Now the system is perturbed by the external potential $\delta\tilde{v}(\vr, t)$ and the thermomechanical potential $\psi(\vr, t)$. The change of the densities to first order is given by 
  \begin{align}
    \begin{pmatrix} \delta n(\vr, t) \\ \delta h(\vr, t) \end{pmatrix}
    & = \ind{}{t'} \ind{3}{r'} \begin{pmatrix} \chi_{nn}(\vr, t; \vrp, t') & \chi_{nh}(\vr, t;\vrp, t') \\
    \chi_{hn}(\vr, t; \vrp, t') & \chi_{hh}(\vr, t;\vrp, t') \end{pmatrix}
    \begin{pmatrix} \delta \tilde{v}(\vrp, t') \\ \delta \psi(\vrp, t') \end{pmatrix} ~. \label{thermalLR}
  \end{align}
  In Eq.\ \eqref{thermalLR} the time integral runs from $-\infty \to \infty$, but the potential is suddenly switched on at $t_0$. Moreover the linear response function $\chi_{AB}$ ($A, B = n, h$) is retarded, i.e., 
  \begin{align}
    \chi_{A B}(t, t') = -i \hbar \theta(t-t')
    \left\langle \left[ \hat{A}(t), \hat{B}(t') \right] \right\rangle ~. \label{LRfunction}
  \end{align}
  Therefore the time integral is only effective from $t_0 \to t$, which means that
  the change in the densities only depends on previous changes in the potential (causality). 
  The time-dependence of the operators in Eq.\ \eqref{LRfunction} implies that the
  operators are taken in the Heisenberg picture w.r.t.\ the unperturbed Hamiltonian, $\hat{H}_0$,
  \begin{align}
    \hat{A}(t) = \exp\left[i \hat{H}_0 t / \hbar\right] \hat{A} \exp\left[-i \hat{H}_0 t / \hbar\right] ~. \label{At}
  \end{align}
  In linear response thermal DFT we are interested in finding the change in the potentials from the 
  variation in the densities. Accordingly we have to investigate under which conditions
  we can invert relation \eqref{thermalLR},
  \begin{align}
    \begin{pmatrix} \delta \tilde{v}(\vr, t) \\ \delta \psi(\vr, t) \end{pmatrix}
    & = \ind{}{t'} \ind{3}{r'} \begin{pmatrix} 
    \big[\chi^{-1}\big]_{nn}(\vr, t; \vrp, t') & \big[\chi^{-1}\big]_{nh}(\vr, t;\vrp, t') \\
    \big[\chi^{-1}\big]_{hn}(\vr, t; \vrp, t') & \big[\chi^{-1}\big]_{hh}(\vr, t;\vrp, t') \end{pmatrix}
    \begin{pmatrix} \delta n(\vrp, t') \\ \delta h(\vrp, t') \end{pmatrix} ~. \label{thermalLRinv}
  \end{align}
  The inverse, $\chi^{-1}$, is only define in the subspace of potentials that induce
  a non-vanishing change in the densities. In order to characterize the kernel we 
  have to search for so-called zero modes, i.e., combinations of the external potential
  perturbation and the variation of the thermomechanical field that do not induce
  changes in the density and the energy density.
  
  Let us recall the Lehmann representation of the response function
  \begin{align}
    \chi_{A B}(t, t') & = -i \hbar \theta(t-t') \sum_{j,k} w_j 
    \Big[ \exp\left[ i \Omega_{jk}(t - t') / \hbar \right] A_{jk} B_{kj}
    - \exp\left[ -i \Omega_{jk}(t - t') / \hbar \right] B_{jk} A_{kj} \Big] \nn
    & = -i \hbar \theta(t-t') \sum_{j,k} (w_j - w_k)
    \exp\left[ i \Omega_{jk}(t - t') / \hbar \right] A_{jk} B_{kj} ~, \label{LRfunctionLehmann}
  \end{align}
  with ${\Omega_{jk} = \epsilon_j - \epsilon_k}$ being the differences of the
  eigenvalues of $\hat{H}_0$, ${A_{jk} = \langle j | \hat{A} | k \rangle}$ are the matrix
  elements of $\hat{A}$ in the basis formed by the eigenstates of $\hat{H}_0$,
  and $w_j$ are the \emph{statistical} weights associated with the energy $\epsilon_j$.
  This implies that we have $w_j \leq w_k$ if $\epsilon_j \geq \epsilon_k$, with the equal
  sign holding only for degenerate eigenstates, which will become important soon.
  Without lost of generality we assume that $t_0 = 0$, the initial time at which the system
  is perturbed sets the origin in time, which implies together with the Heaviside step function
  in the definition of the retarded response function that the time integrals run from $0 \to t$
  in the linear response relation Eq.\ \eqref{thermalLR}. The response integral, hence,
  is a convolution of the retarded response function, which can be deconvoluted by means
  of a Laplace transform,
  \begin{align}
    \tilde{f}(s) = \mathcal{L}\big[f\big](s) = \indll{t}{0}{\infty} e^{-s t} f(t) ~. \label{LaplaceTransform}
  \end{align}
  The zero-modes are characterized by a vanishing density response. Therefore, we can write
  down a necessary condition by convoluting the density response once more with the perturbing potentials, i.e.,
  \begin{align}
    0 = F(T) = \indll{t}{0}{T} \ind{3}{r} \Big(\delta \tilde{v}(\vr, T-t) \delta n(\vr, t)
    + \delta \psi(\vr, T - t) \delta h(\vr, t) \Big) ~. \label{zeroModeNecessary}
  \end{align}
  
  Using the Lehmann representation and the linear response relation Eq.\ \eqref{thermalLR} 
  we can rewrite Eq.\ \eqref{zeroModeNecessary} as
  \begin{align}
    0 & = -i \hbar \sum_{j,k} \frac{w_j}{s - i \Omega_{jk} / \hbar }
    \Big( \tilde{V}_{jk}(s)  + \tilde{\Psi}_{jk}(s) \Big)
    \Big( \tilde{V}_{kj}(s) + \tilde{\Psi}_{kj}(s) \Big) + c.c. \nn
    & = 2 \sum_{j,k} \frac{w_j \Omega_{jk}}{s^2 + (\Omega_{jk} / \hbar)^2 }
    \Big| \tilde{V}_{jk}(s)  + \tilde{\Psi}_{jk}(s) \Big|^2 ~. \label{LR2}
  \end{align}
  where we have introduced 
  \begin{subequations}
    \begin{align}
      \tilde{V}_{jk}(s) & = \mathcal{L}\big[V_{jk}](s) ~,
      & V_{jk}(t) & = \ind{3}{r} \delta \tilde{v}(\vr, t) \langle j | \hat{n}(\vr) | k \rangle ~, \label{Vjk} \\
      \tilde{\Psi}_{jk}(s) & = \mathcal{L}\big[\Psi_{jk}](s) ~,
      & \Psi_{jk}(t) & = \ind{3}{r} \delta \psi(\vr, t) \langle j | \hat{h}(\vr) | k \rangle ~. \label{Psijk}
    \end{align}
  \end{subequations}
  In the second line of Eq.\ \eqref{LR2} we have used that density and energy density are hermitian operators.
  Using the anti-symmetry $\Omega_{jk} = -\Omega_{kj}$ we can rewrite the necessary condition for
  having a zero-mode as
  \begin{align}
    0 & = \sum_{j,k} \frac{(w_j - w_k) \Omega_{jk}}{s^2 + (\Omega_{jk} / \hbar)^2 }
    \Big| \tilde{V}_{jk}(s)  + \tilde{\Psi}_{jk}(s) \Big|^2 \nn
    & = 2 \sum_{j < k} \frac{(w_j - w_k) \Omega_{jk}}{s^2 + (\Omega_{jk} / \hbar)^2 }
    \Big| \tilde{\Delta}_{jk}(s) \Big|^2 ~. \label{LR3}
  \end{align}
  The last line has been obtained by recognizing that the summand remains
  invariant if we swap $j \leftrightarrow k$ and that its diagonal $j=k$ vanishes.
  Furthermore, in Eq.\ \eqref{LR3} we have introduced the abbreviation
  \begin{align}
    \tilde{\Delta}_{jk}(s) = \tilde{V}_{jk}(s)  + \tilde{\Psi}_{jk}(s) ~. \label{Delta}
  \end{align}
  Note that all summands are greater or equal to zero, due to the fact that the
  statistical weights decrease if the energies increase. This is an important piece
  of information for it allows us to impose the condition on each summand separately.
  The leading factor,
  \begin{align}
    \frac{(w_j - w_k) \Omega_{jk}}{s^2 + (\Omega_{jk} / \hbar)^2 } ~, \label{preFactor}
  \end{align}
  vanishes only for degenerate states $j, k$, which leads to the necessary condition
  \begin{align}
    0 = \tilde{\Delta}_{jk}(s) ~, \label{DeltaCondition}
  \end{align}
  for all $j,k$ which do not refer to states in a degenerate subspace. 
  Of course $\Delta_{jk}(t)$ and $\tilde{\Delta}_{jk}(s)$ are zero if
  the perturbations vanish, i.e., $\delta\tilde{v}(\vr, t) = \delta \psi(\vr, t) = 0$.
  However, this is the trivial zero-mode and we are interested in non-trivial solutions
  to Eq.\ \eqref{DeltaCondition}.
  
  Non-trivial solution to condition \eqref{DeltaCondition}
  are given by perturbations
  \begin{align}
    \hat{\Delta}_0(t) = \ind{3}{r} \Big( \delta\tilde{v}_0(\vr,t) \hat{n}(\vr)
    + \delta\psi_0(\vr,t) \hat{h}(\vr) \Big) ~, \label{zeroModePerturbation}
  \end{align}
  which satisfy
  \begin{align}
    \hat{\Delta}_0(t) | j \rangle = \sum_{k \in d(j)} \big[\Delta_0\big]_{kj}(t) | k \rangle ~, 
    \label{Delta0}
  \end{align}
  where $d(j)$ denotes the space of eigenstates of $\hat{H}_0$ which have the same eigenvalue
  as state $j$. Equation \eqref{Delta0} means that the perturbation only mixes
  degenerate eigenstates. Furthermore, Eq.\ \eqref{Delta0} has to be true for all eigenstates $k$
  since in a statistical ensemble all states have a non-vanishing weight $w_j$. Since the perturbation
  then only mixes degenerate subspaces of $\hat{H}_0$ this implies
  \begin{align}
    0 = \Big[ \hat{\Delta}_0, \hat{H}_0 \Big] ~, \label{SymmetryCondition}
  \end{align}
  which means that $\hat{\Delta}_0$ represents a symmetry of the system, i.e., it acts like a ``phantom perturbation".
  
  So far we have only established a necessary condition for having a vanishing response.
  Now, we confirm that this condition is indeed sufficient. In order to see this explicitly
  we write out the density variations using the Lehmann representation
  Eq.\ \eqref{LRfunctionLehmann},
  \begin{subequations} 
    \begin{align}
      \delta n(\vr, t) & = -i \hbar \indll{t'}{0}{t} \sum_{j,k} (w_j - w_k)
      \exp\left[ i \Omega_{jk}(t - t') / \hbar \right] \langle j | \hat{n}(\vr) | k \rangle
      \Big( V_{kj}(t') + \Psi_{kj}(t') \Big) ~, \label{dns1} \\
      \delta h(\vr, t) & = -i \hbar \indll{t'}{0}{t} \sum_{j,k} (w_j - w_k)
      \exp\left[ i \Omega_{jk}(t - t') / \hbar \right] \langle j | \hat{h}(\vr) | k \rangle
      \Big( V_{kj}(t') + \Psi_{kj}(t') \Big) ~. \label{dhs1}
    \end{align}
  \end{subequations}
  Now we plug in the potentials that fulfill the necessary condition for having a
  zero-mode. From Eq.\ \eqref{Delta0} we get
  \begin{subequations} 
    \begin{align}
      \delta n(\vr, t) & = -i \hbar \indll{t'}{0}{t} \sum_{j,k \in d(j)} (w_j - w_k)
      \exp\left[ i \Omega_{jk}(t - t') / \hbar \right] \langle j | \hat{n}(\vr) | k \rangle
      \big[\Delta_0\big]_{kj}(t') = 0 ~, \label{dns2} \\
      \delta h(\vr, t) & = -i \hbar \indll{t'}{0}{t} \sum_{j,k \in d(j)} (w_j - w_k)
      \exp\left[ i \Omega_{jk}(t - t') / \hbar \right] \langle j | \hat{h}(\vr) | k \rangle
      \big[\Delta_0\big]_{kj}(t') = 0 ~, \label{dhs2}
    \end{align}
  \end{subequations}
  which shows that the necessary condition is also sufficient. The crucial observation
  is that the statistical weights are identical if $j$ and $k$ are in the same degenerate
  subspace of the Hamiltonian $\hat{H}_0$.
  
  In conclusion we have shown that the thermal response function is invertible if we exclude
  perturbations which represent symmetries of the Hamiltonian. If we focus on continuous symmetries
  we only have to check if the generators of the symmetry group can be written
  as superpositions of the density and the energy density. Obvious generators are the
  unperturbed Hamiltonian itself, as generator for translations in time, and the
  total particle number operator, as generator of total phase shifts. We point out that
  the generators for other common symmetries, such as translation and rotations in space,
  cannot be written in terms of the density and energy density. Even though it cannot
  be categorically excluded that there are other symmetries, generated by a combination
  of the density and energy density, it appears that they would be rather unusual symmetries.
\end{appendix}

\bibliography{ThermoelectricityReview}

\end{document}